\documentclass[superscriptaddress,nobibnotes,aps,twocolumn,prd,showpacs,nofootinbib]{revtex4-1}
\usepackage{amsmath}

\DeclareMathOperator{\sech}{sech}
\usepackage{amsfonts}
\usepackage{amssymb}
\usepackage{graphicx}
\makeatletter
\newcommand*{\rom}[1]{\expandafter\@slowromancap\romannumeral #1@}
\makeatother

\usepackage[usenames,dvipsnames]{xcolor}
\usepackage{hyperref}
\hypersetup{colorlinks=true,
citecolor=MidnightBlue, linkcolor=MidnightBlue, urlcolor=MidnightBlue}
\usepackage{tikz}
\definecolor{purple}{rgb}{1,0,1}
\definecolor{lime}{HTML}{A6CE39} % needs xcolor

\usepackage{epsfig}

\newcommand{\orcidicon}{%
	\begin{tikzpicture}
	\draw[lime, fill=lime] (0,0)
		circle [radius=0.16]
		node[white] {{\fontfamily{qag}\selectfont \tiny ID}};
	\draw[white, fill=white] (-0.0625,0.095)
		circle [radius=0.007];
	\end{tikzpicture}	\hspace{-2mm}
}

\newcommand\orcidFrancisco{{\href{https://orcid.org/0000-0002-9388-8373}{\orcidicon}}}
\newcommand\orcidMarziyeh{{\href{https://orcid.org/0000-0003-2736-9396}{\orcidicon}}}
\begin{document}
%%%%%%%%%%%%%%%%%%%%%%%%%%%%%%%%%%%%%%%%%%%%%%%%%%%%

%%%%%%%%%%%%%%%%%%%%%%%%%%%%%%%%%%%%%%%%%%%%%%%%%%%%
%\title{ The effect of symmetry breaking on the structure of thick brane}
\title{Novel thick brane solutions with $U(1)$ symmetry breaking}
%%%%%%%%%%%%%%%%%%%%%%%%%%%%%%%%%%%%%%%%%%%%%%%%%%%%
\author{Marzieh Peyravi\orcidMarziyeh\!\!}
\email{marziyeh.peyravi@stu-mail.um.ac.ir}
\affiliation{Department of Physics, School of Sciences, Ferdowsi University
of Mashhad, Mashhad 91775-1436, Iran}

\author{Nematollah Riazi}
\email{n\_riazi@sbu.ac.ir}
\affiliation{Physics Department, Shahid Beheshti University, Evin, Tehran 19839, Iran}

\author{Francisco S. N. Lobo\orcidFrancisco\!\!}
\email{fslobo@fc.ul.pt}
\affiliation{Instituto de Astrof\'{\i}sica e Ci\^{e}ncias do Espa\c{c}o, Faculdade de
Ci\^encias da Universidade de Lisboa, Edif\'{\i}cio C8, Campo Grande,
P-1749-016 Lisbon, Portugal}
%%%%%%%%%%%%%%%%%%%%%%%%%%%%%%%%%%%%%%%%%%%%%%%%%%%%

%%%%%%%%%%%%%%%%%%%%%%%%%%%%%%%%%%%%%%%%%%%%%%%%%%%%
\begin{abstract}
%%%%%%%%%%%%%%%%%%%%%%%%%%%%%%%%%%%%%%%%%%%%%%%%%%%%

In this work, using two scalar fields ($\phi$, $\psi$) coupled to 4+1 dimensional gravity, we construct novel topological brane solutions through an explicit $U(1)$ symmetry breaking term. The potential of this model is constructed so that two distinct degenerate vacua in  the $\phi$ field exist, in analogy to the $\phi^{4}$ potential. Therefore, brane solutions  appear due to the vacuum structure of the $\phi$ field. However, the topology and vacuum structure in the $\psi$ direction depends on the symmetry breaking parameter $\beta^{2}$, which leads to different types of branes. As a result, one can interpret the present model as a combination of a $\phi^{4}$ brane with an auxiliary field, which leads to deviations from the $\phi^{4}$ system with the brane achieving a richer internal structure. Furthermore, we analyse in detail the behaviour of the superpotentials, the warp factors, the Ricci and Kretschmann scalars and the Einstein tensor components. In addition to this, we explore the stability of the brane in terms of the free parameters of the model. The analysis presented here complements previous work and is sufficiently novel to be interesting.\\


\end{abstract}
%%%%%%%%%%%%%%%%%%%%%%%%%%%%%%%%%%%%%%%%%%%%%%%%%%%%

%%%%%%%%%%%%%%%%%%%%%%%%%%%%%%%%%%%%%%%%%%%%%%%%%%%%
\date{\today }
\maketitle
%%%%%%%%%%%%%%%%%%%%%%%%%%%%%%%%%%%%%%%%%%%%%%%%%%%%

%%%%%%%%%%%%%%%%%%%%%%%%%%%%%%%%%%%%%%%%%%%%%%%%%%%%
%\tableofcontents
%%%%%%%%%%%%%%%%%%%%%%%%%%%%%%%%%%%%%%%%%%%%%%%%%%%%

%%%%%%%%%%%%%%%%%%%%%%%%%%%%%%%%%%%%%%%%%%%%%%%%%%%%
\section{Introduction}
%%%%%%%%%%%%%%%%%%%%%%%%%%%%%%%%%%%%%%%%%%%%%%%%%%%%

The braneworld scenario describes our $4$-dimensional observable universe as a localized brane embedded in a $4+d$-dimensional spacetime, denoted the ``bulk'', with Standard Model particles and fields trapped on the brane while gravity is free to propagate in the bulk \cite{Randall:1999ee,Randall:1999vf,Dvali:2000hr,Shiromizu:1999wj,Maartens:2003tw}. In this context, scalar fields can generate topological structures even in the absence of gravity, and thus induce localized brane scenarios \cite{DeWolfe:1999cp}.
In fact, different types of localized structures exist, such as domain walls, strings, monopoles and vortexes  \cite{Mukhanov:2005sc,Vilenkin:2000jqa, Manton:2004tk,Dzhunushaliev:2009va}. The exact shape of the structures depend essentially on the physics, spacetime dimensions, symmetry breaking and the topology of the vacua.
Based on the literature, at least three classes of models exist that support these kink-like defects \cite{Bazeia:2015eta,Bazeia:2004dh,Sadeghi:2007uk,Afonso:2006gi,Bazeia:2002xg,Cruz:2014eca,Bazeia:2004dh,Mannheim:2005br,Flanagan:2001dy,Peyravi:2015bra,Veras:2015lyz}. The first class deals with a single real scalar field, which leads to structureless topological solutions, such as the sine-Gordon and $\phi^{4}$ models. In soliton theory, the latter possess simple soliton-like solutions, and have been the object of building thick branes \cite{Zeldovich:1974uw,Vilenkin:1981zs,Vilenkin:1984hy,Vilenkin:1984ib}. The second class contains a single real scalar field, but now the system admits at least two distinct types of branes (walls), as for instance in the double sine-Gordon model. The third class is defined by two real scalar fields, which essentially induce an internal brane structure \cite{Dzhunushaliev:2009va,Bazeia:2004dh}.

In many brane world scenarios, only one scalar field is responsible for generating the brane \cite{Peyravi:2015bra,Mannheim:2005br},
however, inspired by condensed matter physics and ferromagnetic systems, an Ising or Bloch-type domain wall has been considered as a brane candidate \cite{deSouzaDutra:2008gm,Bazeia:2004dh,Dzhunushaliev:2009va}. More specifically, an Ising wall is a simple interface without an internal structure, while the Bloch version is an interface which has a nontrivial internal structure and so possesses features that are not present in the case of a single field. Thus, it is useful to explore Bloch walls as thick brane solutions with an internal structure \cite{Bazeia:2004dh, Dzhunushaliev:2009va}.
Indeed, the Bloch brane models are constructed based on the interaction of two real scalar fields coupled with gravity in $4+1$ dimensional warped spacetime involving one extra dimension. The field interaction depends on a real parameter which determines the way the scalar field interacts with itself, which is a generalized form of the standard $\phi^{4}$ model. In fact, Bloch branes have more localized solutions and as a result, a much richer structure, which is specified by degeneracy controlling parameters \cite{deSouzaDutra:2008gm}.

The literature has extensively explored thick branes with internal structure induced by the parameter that controls the interaction between two scalar fields coupled to gravity in 4+1 dimensions. For instance, in \cite{Bazeia:2002xg} a general method, valid for both topological and non-topological defects, was introduced to obtain deformed defects starting from a given scalar field theory. The procedure allowed the construction of infinitely many new theories that support defect solutions, which were analytically expressed in terms of the defects of the original theory. In this manner, without changing the corresponding topological behavior, one can vary the amplitude and width of the domain wall \cite{Bazeia:2002xg}. In \cite{deSouzaDutra:2008gm} it was also shown that one may control the thickness of the domain walls by an external parameter without changing the parameters of the potential.
The discovery of stable multikink solutions in thick brane models, that move with large velocities, has also motivated the development of brane scenarios with double and multi-brane configurations with symmetric and asymmetric warp factors in order to solve the hierarchy problem in thick brane scenarios
\cite{Dutra:2014xla,Melfo:2002wd,Melfo:2006hh,Liu:2009dw,Zhao:2009ja,Cruz:2011ru,Chumbes:2011zt,Xie:2013rka,Cruz:2013zka,Bazeia:2013usa,deSouzaDutra:2008gm,Ahmed:2012nh,Dutra:2013jea,Peyrard:1985uf,Dzhunushaliev:2008zz}.

In this work, we are interested in exploring the effect of an explicit symmetry breaking by considering two coupled fields ($\phi$, $\psi$), with a potential that is similar to that of the hybrid inflation potential \cite{Linde:1993cn}, which has well-known solutions in 5D spacetime. The potential of this model has been constructed so that two distinct degenerate vacua in  the $\phi$ field exist, in analogy to the $\phi^{4}$ system potential. Therefore, brane solutions  appear due to the vacuum structure of the $\phi$ field. However, the topology and vacuum structure in the $\psi$ direction depends on the symmetry breaking parameter $\beta^2$, which will lead to two different types of branes. As a result, one can interpret the present model as a combination of a $\phi^{4}$ brane with an auxiliary field, which leads to deviations from the $\phi^{4}$ system with the  brane achieving a richer internal structure. In other words, the second field is significant because one can control and modify the configuration of the $\phi^{4}$ field and the brane by this extra field.

The paper is outlined in the following manner: In Sec. \ref{11}, we outline the general formalism of the brane world scenario, by writing the action and gravitational field equations, and analyse the particle motion near the brane through the geodesic equations. In  Sec. \ref{22}, we present novel thick brane solutions with $U(1)$ symmetry breaking, by specifying the double field potential, and present the soliton solutions. Furthermore, we analyse in detail the behaviour of the superpotentials, the warp factors, the Ricci and Kretschmann scalars and the mixed Einstein tensor components, in addition to exploring the stability regions of the potential of the linearized Schr\"{o}dinger equation as a function of the free parameters of the model. Finally, in Sec. \ref{44}, we conclude.

%%%%%%%%%%%%%%%%%%%%%%%%%%%%%%%%%%%%%%%%%%%%%%%%%%%%
\section{5-D thick brane: General formalism}\label{11}
%%%%%%%%%%%%%%%%%%%%%%%%%%%%%%%%%%%%%%%%%%%%%%%%%%%%

%%%%%%%%%%%%%%%%%%%%%%%%%%%%%%%%%%%%%%%%%%%%%%%%%%%%
\subsection{Action and field equations}
%%%%%%%%%%%%%%%%%%%%%%%%%%%%%%%%%%%%%%%%%%%%%%%%%%%%

We consider a thick brane, embedded in a five-dimensional (5D) bulk spacetime, modelled by the following action \cite{deSouzaDutra:2008gm,Bazeia:2004dh,Dzhunushaliev:2009va}:
\begin{eqnarray}
S=\int
d^{5}x\sqrt{|g^{(5)}|}\left[-\frac{1}{4} R[g^{(5)}]+\frac{1}{2}\partial_{B}\phi\partial^{B}\phi
	\right.
	\nonumber  \\
	\left.
+\frac{1}{2}\partial_{B}\psi\partial^{B}\psi-V(\phi,\psi)\right],
\label{actionbrane}
\end{eqnarray}
where $g^{(5)}$ is the metric and $R[g^{(5)}]$ the scalar curvature in the bulk;
$\phi$ and $\psi$ are dilaton fields living in the bulk and $V(\phi,\psi)$ is a general potential energy; we have used $\kappa_{5}^{2}=8\pi G_{5}=2$.
It is interesting to note that historically the action (\ref{actionbrane}), with the potential (\ref{mpot}),
and the superpotential method given below, was considered in the context of higher dimensional supergravity theories, in the 1980s \cite{Boucher:1984yx,Townsend:1984iu}.

The simplest line element of the brane, embedded in the 5D bulk spacetime with metric signature $(+,-,-,-,-)$ can be written as \cite{deSouzaDutra:2008gm,Bazeia:2004dh,Dzhunushaliev:2009va}:
\begin{eqnarray}
ds^{2}_{5}&=&g_{CD}dx^{C}dx^{D}\nonumber\\
&=&e^{2A}\eta_{\mu\nu}dx^{\mu}dx^{\nu}-dw^{2},
\end{eqnarray}
where $C, D=1...5$, $\mu, \nu=1...4$ and $A$ is the warp function which depends only on the fifth coordinate $w$. The 5D energy-momentum tensor of the system is given by:
\begin{eqnarray}
&&T_{CD}=\partial_{C}\phi\partial_{D}\phi+\partial_{C}\psi\partial_{D}\psi
	\nonumber  \\
&&\quad -g_{CD}\left[\frac{1}{2}\partial_{B}\phi\partial^{B}\phi
+\frac{1}{2}\partial_{B}\psi\partial^{B}\psi-V(\phi,\psi)\right],
\end{eqnarray}
where the metric functions $g_{CD}$, and the scalar fields, $\phi$  and $\psi$, depend solely on $w$. Note that we have ignored the standard model matter on the brane at this stage, and the sole source of the energy-momentum is the two scalar fields.

The 5D gravitational field and the equations of motion for the scalar fields take the following forms \cite{deSouzaDutra:2008gm,DeWolfe:1999cp,Bazeia:2004dh,Dzhunushaliev:2009va}:
\begin{eqnarray}
A^{\prime\prime}&=&-\frac{2}{3}\left[{\phi^{\prime}}^{2}+{\psi^{\prime}}^{2}\right] ,
\label{4567}
     \\
{A^{\prime}}^{2}&=&\frac{1}{6}\left[{\phi^{\prime}}^{2}+{\psi^{\prime}}^{2}\right]-\frac{1}{3}V(\phi,\psi) ,
\label{4567b}
     \\
\phi^{\prime\prime}+4A^{\prime}\phi^{\prime}&=&\frac{\partial
V(\phi,\psi)}{\partial\phi},
\label{4567c}
     \\
\psi^{\prime\prime}+4A^{\prime}\psi^{\prime}&=&\frac{\partial
V(\phi,\psi)}{\partial\psi},
\label{4567d}
\end{eqnarray}
respectively, where the prime denotes a derivative with respect to $w$.

In order to replace the second order differential equations (\ref{4567c}) and (\ref{4567d}) with first order equations, it is useful to introduce a superpotential $W(\phi,\psi$) \cite{deSouzaDutra:2008gm,DeWolfe:1999cp,Bazeia:2004dh,Dzhunushaliev:2009va,Sadeghi:2007uk}, which demands:
\begin{eqnarray}
A^{\prime}&=&-\frac{1}{3}W(\phi,\psi),\label{8910} \\
\phi^{\prime}&=&\frac{1}{2}\frac{\partial
W(\phi,\psi)}{\partial\phi}, \label{8910b} \\
\psi^{\prime}&=&\frac{1}{2}\frac{\partial
W(\phi,\psi)}{\partial\psi},\label{8910c}
\end{eqnarray}
while $V(\phi,\psi)$ takes the following form \cite{deSouzaDutra:2008gm,DeWolfe:1999cp,Bazeia:2004dh,Dzhunushaliev:2009va,Sadeghi:2007uk}:
\begin{eqnarray}\label{mpot}
V(\phi,\psi)&=&\frac{1}{8}\left[\left(\frac{\partial
W(\phi,\psi)}{\partial\phi}\right)^{2}+\left(\frac{\partial
W(\phi,\psi)}{\partial\psi}\right)^{2}\right]
	\nonumber \\
&&\qquad -\frac{1}{3}W(\phi,\psi)^{2}.
\end{eqnarray}
We note that the corresponding first order equations for solution of the type analysed in this paper were first given in the context of dilaton domain walls of D-dimensional gravity with the general dilaton potential admitting a stable anti-de Sitter vacuum \cite{Skenderis:1999mm}.

Now, to solve Eqs. (\ref{4567})--(\ref{4567d}) for thick brane solutions, there exist two different approaches. In the first approach, one starts with presumed exact static solutions of fields and determines the rest of variables such that all equations are satisfied consistently. More specifically, one starts from solitonic solutions in flat spacetime and then by solving the nonlinear equations one must modify the scalar
field potential in such a way that the soliton solutions remain
a solution of the full gravitating system. Then the superpotential $W(\phi,\psi)$ and warp function would be calculated. In this method the soliton solution remains the same in flat and curved spacetime, although the form of the potential, changes accordingly \cite{Mannheim:2005br,DeWolfe:1999cp}.

In the second approach one starts from a specific superpotential, instead of the fields ($\phi$ and $\psi$). This approach is based on minimizing the energy (Bogomolny bound) and imposing parity restrictions (for instance, see \cite{Bazeia:2004dh}). Thus, one should solve the following equation in order to check whether solitons exist or not and which kind of solitons will appear in the system \cite{deSouzaDutra:2008gm,Bazeia:2004dh,Dzhunushaliev:2009va}:
\begin{equation}\label{grant}
\frac{d\phi}{d\psi}=\frac{W_{\phi}}{W_{\psi}}=\frac{\phi^{\prime}}{\psi^{\prime}},
\end{equation}
which is the general nonlinear differential equation relating the scalar fields of the model \cite{deSouzaDutra:2008gm,DeWolfe:1999cp,Bazeia:2004dh,Dzhunushaliev:2009va,Bazeia:1995en}.
If solutions exist in the form of $\phi(\psi)$ \cite{deSouzaDutra:2008gm}, this function represents the equation for a generic orbit, which reflects the presence of topological soliton solutions \cite{Rajaraman:1978kd,Bazeia:1995en}. We are interested in potentials which lead to orbits in the ($\phi$, $\psi$) plane corresponding to topological solitons, namely, potentials with a unique minimum in $\psi$ and two degenerate minima in the $\phi$ direction which lead to topological solitons. One can show that such orbits will have zero constant of integration ($\frac{1}{2}\phi'^{2}+\frac{1}{2}\psi'^{2}+\tilde{V}(\phi,\psi)=0$) and finite total energy ($\int_{-\infty}^{+\infty}dw \left(\frac{1}{2}\phi'^{2}+\frac{1}{2}\psi'^{2}-\tilde{V}(\phi,\psi)\right)={\rm const}$). These results from soliton theory are very helpful in constructing thick brane models presented in this paper.

The energy density distribution on the bulk, $T_{00}$, which will be analyzed in detail below, is given by:
\begin{equation}
T_{00}=e^{2A}\left[\frac{1}{2}\left(\frac{\partial\phi}{\partial
w}\right)^{2}+\frac{1}{2}\left(\frac{\partial\psi}{\partial
w}\right)^{2}+V(\phi,\psi)\right].
\label{energydensity_bloch}
\end{equation}
It can also be shown that for models with an infinitely thin brane and Dirac delta distributions, the energy density is equal to the cosmological constant of the bulk ($\Lambda_{5}^{\pm})$ plus the energy density on the brane, i.e., $\varepsilon=\Lambda_{5}^{\pm}+k\delta(w)$ \cite{Peyravi:2015bra}, where $k$ is parameter independent of $w$ and is related to the energy density on the brane.

%%%%%%%%%%%%%%%%%%%%%%%%%%%%%%%%%%%%%%%%%%%%%%%%%%%%
\subsection{Geodesic equation}
%%%%%%%%%%%%%%%%%%%%%%%%%%%%%%%%%%%%%%%%%%%%%%%%%%%%

Moreover, it is also interesting to investigate the particle motion near the brane \cite{Jalalzadeh:2004uv} through the geodesic equation along the fifth dimension in a thick brane. This investigation helps in clarifying the interaction of material particles to the gravitational field of the brane, in particular, whether matter is gravitationally confined to the brane. To this effect, the geodesic equation provides one with the differential equations:
\begin{equation}
\frac{d}{d\tau}\left(-2e^{2A}\dot{t}\right)=0, \qquad \ddot{w}+A'e^{2A}\dot{t}^{2}=0,
\end{equation}
which yield
\begin{equation}\label{geo}
\ddot{w}-c_{1}^{2}\left[f(w)\right]=0 ,
\end{equation}
where the overdot denotes a derivative with respect to the proper time $\tau$ on the brane, $c_{1}$ is a constant of integration and the factor $f(w)$ is defined as
\begin{equation}
f(w)=A'(w)e^{-2A(w)}.
\end{equation}

Equation (\ref{geo}) is a  second order differential equation for $w$ and its solution depends critically on whether $f(w)$ is positive or negative near to the position of the brane, i.e., $w\approx0$. For positive (negative) values of $f(w)$ one obtains exponential (periodic) solutions, respectively. Note that the periodic motion indicates particle confinement near the brane, while the exponential solutions implies that the reference point is unstable. However, this may show that the reference point is chosen incorrectly and the brane is located at $w\neq 0$ rather than $w=0$.

In a periodic situation, by introducing a new quantity
$F(w)=-c_{1}^{2}A'(w)e^{-2A(w)}$, one can write the geodesic equation in the following form
\begin{equation}
\ddot{w}+F(w)=0.
\end{equation}
One can show that in the exact stable point, i.e., $w_{0}$,  we have $F(w_{0})=0$. On the other hand, by expanding $F(w)$ around $w_{0}$, we have $F(w)=F(w_{0})+F'(w_{0})(w-w_{0})+...$,
%\begin{equation}
%F(w)=F(w_{0})+F'(w_{0})(w-w_{0})+... \,,
%\end{equation}
and the geodesic equation leads to $\ddot{w}+F'(w_{0})(w-w_{0})=0.$
%\begin{equation}
%\ddot{w}+F'(w_{0})(w-w_{0})=0.
%\end{equation}
Taking into account a change of variable $\tilde{w}=w-w_{0}$, the geodesic equation reduces to
\begin{eqnarray}
\ddot{\tilde{w}}+\Omega^{2}\,\tilde{w} = 0 \,,
\end{eqnarray}
where $\Omega=\sqrt{F'(w_{0})}$. Note that the stability of the orbits and the gravitational confinement of particles to the brane requires $F'(w_{0}) \geq 0 $.

One may interpret these results in that the thick branes, via the scalar fields $\phi$ and $\psi$, provides a gravitational field which confines test particles to the vicinity of the brane, forcing them to oscillate along $w$ between either sides. For thin branes, matter particles are strictly confined to the brane via a delta function $\delta(w)$ which appears in the energy-momentum tensor. In thick branes, on the other hand, particles are strongly attracted toward the brane location via a confinement mechanism. For this reason, we consider particle motion and confinement in the vicinity of the brane.

%%%%%%%%%%%%%%%%%%%%%%%%%%%%%%%%%%%%%%%%%%%%%%%%%%%%
\section{Thick Branes with $U(1)$ symmetry breaking}\label{22}
%%%%%%%%%%%%%%%%%%%%%%%%%%%%%%%%%%%%%%%%%%%%%%%%%%%%

\begin{figure*}[ht!]
\epsfxsize=9cm\centerline{\hspace{8cm}\epsfbox{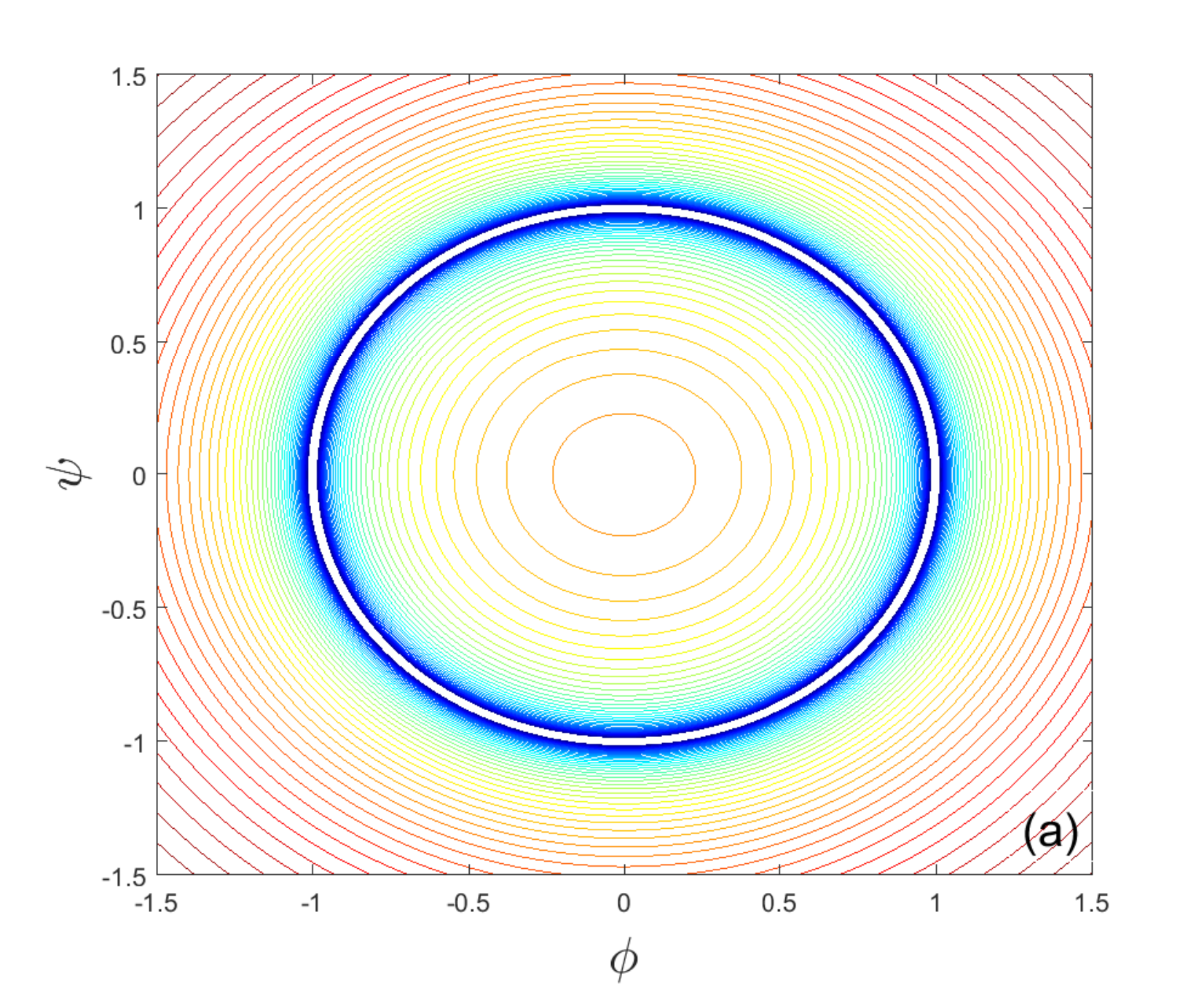}\epsfxsize=9cm\centerline{\hspace{-10.2cm}\epsfbox{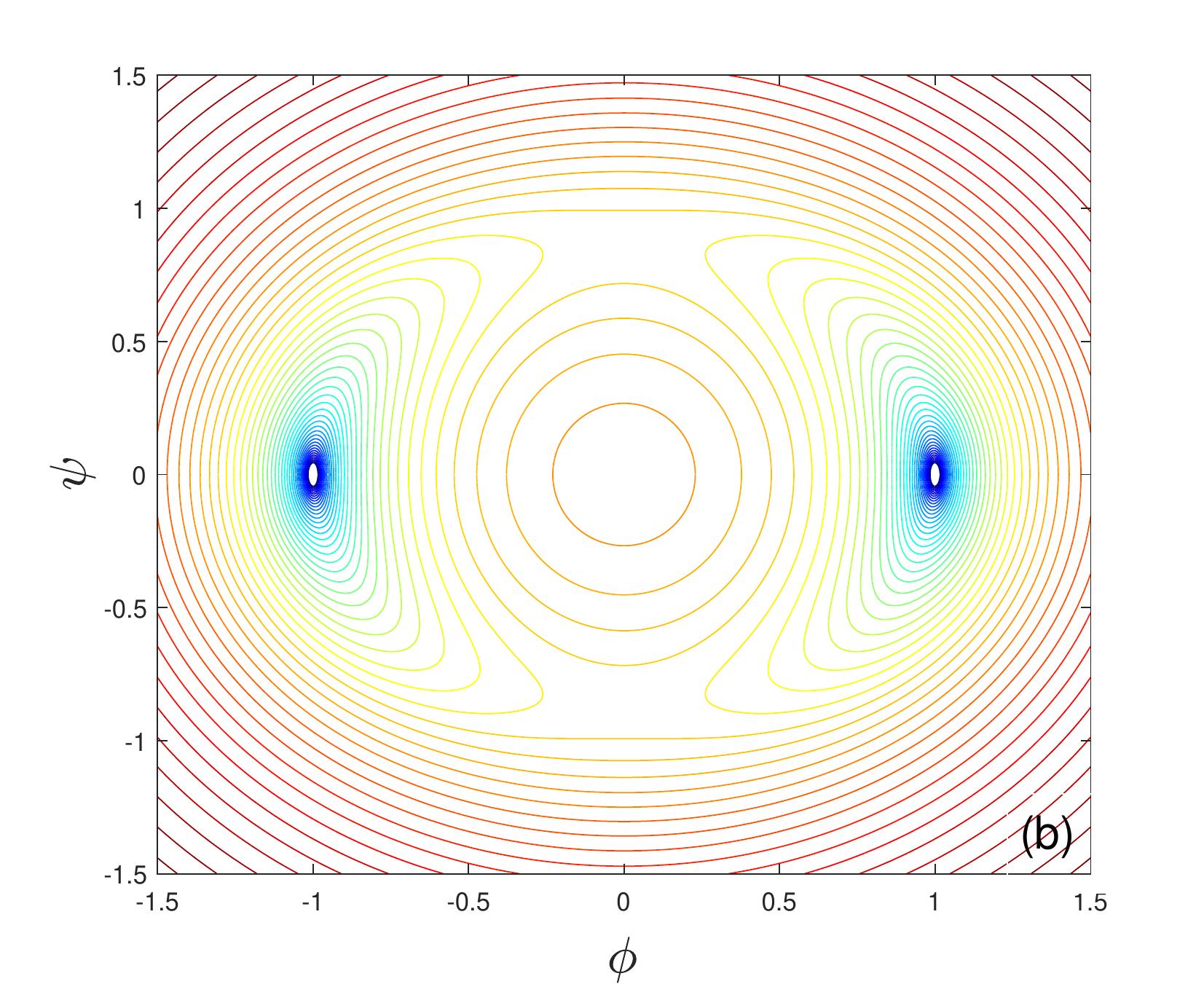}}}
\caption{The logarithm of the potential $V(\phi$, $\psi$) is depicted as equipotential contours on the ($\phi$, $\psi$) plane. We have considered the logarithmic scale in order to show the behaviour of the surface curves as they vary from maximum to minimum points, which are indicated by the color spectrum, from red to blue, respectively. The potential increases from blue to red. (a) $\lambda=1$, $\alpha_{\phi}=1$ and $\beta^{2}=0$ and (b) $\lambda=1$, $\alpha_{\phi}=1$ and  $\beta^{2}=1$. It can be seen that the vacuum is $S^{1}$ for $\beta^{2}=0$ and a discrete set of two points for $\beta^{2}=1$ (right panel).}
\label{2}
\end{figure*}

\begin{figure*}[ht!]
\epsfxsize=16.0cm\centerline{\epsfbox{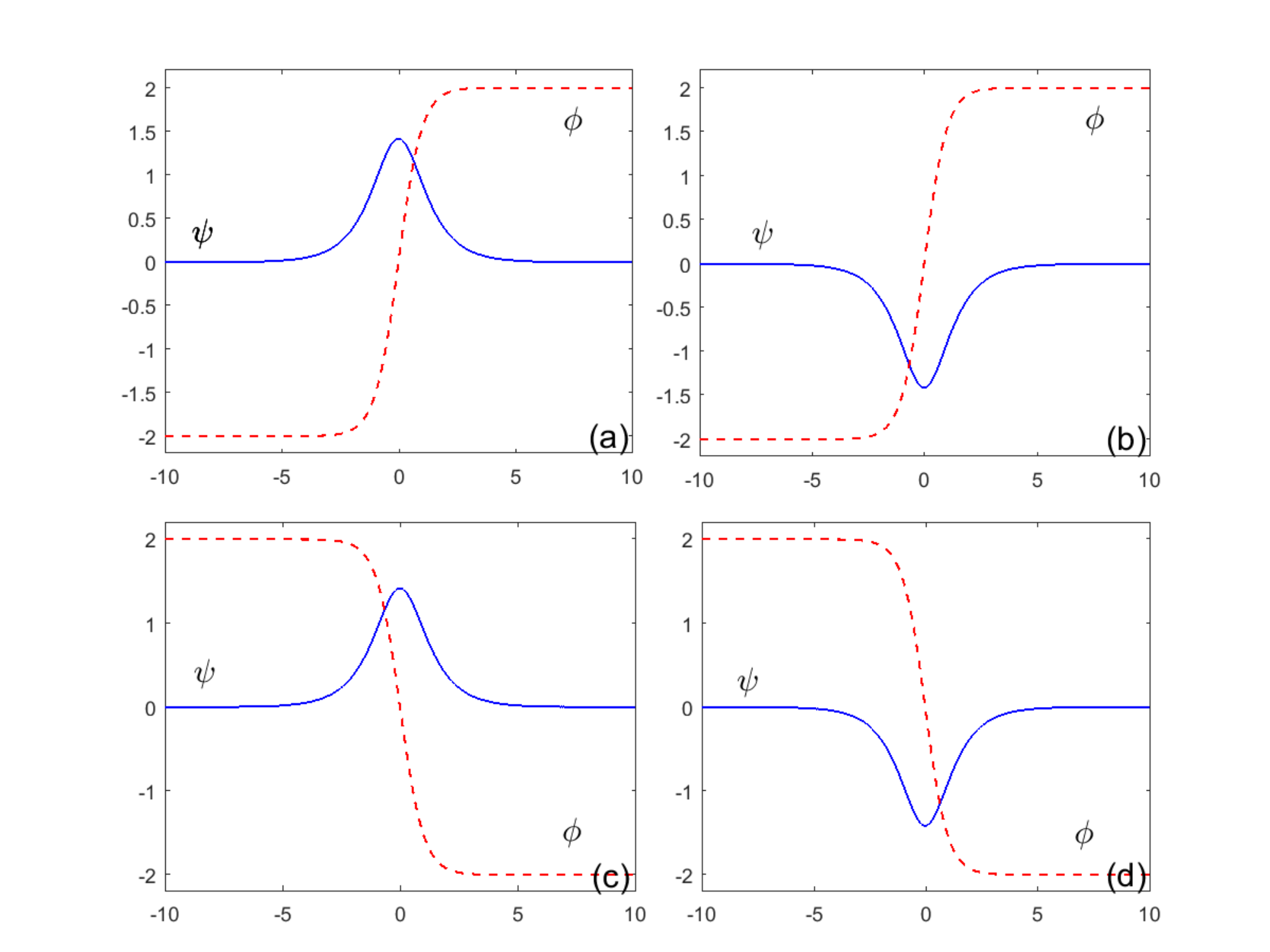}} \caption{Static solutions of Eq. (\ref{kak}). The dashed curves represent $\phi$ and the solid curves are for $\psi$, for $\lambda=1$, $\alpha_{\phi}=2$, $\alpha_{\psi}=1.4142$ and $\beta=1$. As one can see, under the transformation $(\phi,\psi)\rightarrow
(-\phi, -\psi)$, the soliton $k_1$ (Fig. \ref{3}(a))
changes to $\tilde{k}_2$ (Fig. \ref{3}(d))  and $\tilde{k}_1$ (Fig. \ref{3}(c))  changes to $k_2$ (Fig. \ref{3}(b)).
\label{3}}
\end{figure*}

We consider, as is predicted in thick brane scenarios, the observable universe to be a brane-like structure in a 5-dimensional bulk spacetime. The brane is located somewhere along the extra dimension $w$, say $w=w_{0}$. In thick brane models, there is no need to apply the junction conditions at $w=w_{0}$, since the metric and the matter distribution vary smoothly along $w$.
Thick branes can be divided into two distinct classes: topological and nontoplogical. In topological thick branes, the scalar field(s) which support the brane rest on two \emph{distinct} values, corresponding to distinct degenerate vacua of the field, while in nontopological branes this is not the case.
The topological thick branes could also contain a simple field ($\phi$), or multiple fields ($\phi_{a}=1,2,..N$), depending on the number of independent scalar fields which appear in the Lagrangian.

When there are $n$ extra dimensions and $n$ scalar fields with $S_{n}$ topology for the degenerate vacua, for instance, as in the $V(\phi)=\frac{\lambda}{4}\left(\phi^{a}\phi^{a}-\eta^{2}\right)^{2}$ model, different types of solutions are found \cite{Dzhunushaliev:2009va}.
Such highly symmetrical branes, although interesting theoretically, are idealized models. In fact, what we learn from particle and condensed matter physics is that most symmetries are broken  spontaneously, explicitly, or broken at the quantum level \cite{Guidry}. Our main motivation in the present work is to explore the effect of a symmetry breaking term in an otherwise $U(1)$ symmetric Lagrangian. This extra term turns out to determine the internal structure and normal modes of the brane solutions.

%%%%%%%%%%%%%%%%%%%%%%%%%%%%%%%%%%%%%%%%%%%%%%%%%%%%
\subsection{Specific double field potential}
%%%%%%%%%%%%%%%%%%%%%%%%%%%%%%%%%%%%%%%%%%%%%%%%%%%%

The double field potential we are interested in is given by \cite{Assyyaee:2015ale}:
\begin{eqnarray}\label{a}
\tilde{V}(\phi,\psi)=\frac{\lambda}{4}\left(\phi^2+\psi^2-\alpha_{\phi}^{2}\right)^2+\frac{1}{2}\beta^{2}\psi^2,
\end{eqnarray}
where  $\alpha_{\phi}$ is a constant parameter which controls the height the brane, $\beta^{2}$ is a constant parameter which controls the thickness of the brane as well as controlling the $U(1)$ explicit symmetry breaking and $\phi$ and $\psi$ are real scalar fields. The term within the parenthesis of this potential demonstrates a full circular symmetry resembling the Higgs potential \cite{Assyyaee:2015ale}. This potential is similar to, but not the same as, that of the hybrid inflationary model
\cite{Assyyaee:2015ale,paper4,Linde:1993cn}. In the latter, there are two scalar fields, one playing the role of a rapidly decaying (water-fall) field, triggered by another (inflationary) scalar field \cite{Assyyaee:2015ale,paper4,Linde:1993cn}. In such models, depending on the choice of the Lagrangian density, the model may lead to the formation of domain walls in $3 + 1$ dimensions.
As depicted in Fig. \ref{2}, the cases $\beta^{2}=0$ and $\beta \neq 0$ correspond to two different
topologies for the vacuum and as a result distinct topological solitons. One can demonstrate for $\beta^{2}=0$, that the Lagrangian density will be Lorentz invariant as well as
invariant under a global $U(1)$ transformation. However, for $\beta^{2}\neq0$ the $U(1)$ symmetry is broken and the potential along the $\phi$ axis has always two degenerate vacua at $\phi=\pm \alpha$, while the potential along the $\psi$ axis depend on $\beta^{2}$.

One can show that for $\lambda\alpha_{\phi}^{2}>\beta^{2}$ the potential has two saddle points which are located at $\psi=\pm\sqrt{\alpha_{\phi}^{2}-\beta^{2}/\lambda}$
on the $\psi$ axis which are transformed to each other by a sign transformation \cite{Assyyaee:2015ale,paper4}. It is worthwhile to emphasize that the saddle points move toward the origin as the inequality $\lambda\alpha_{\phi}^{2}>\beta^{2}$
 becomes weaker and finally meet at the origin when $\lambda\alpha_{\phi}^{2}=\beta^{2}$ \cite{Assyyaee:2015ale}. However for $\lambda\alpha_{\phi}^{2}\leqslant\beta^{2}$, the origin remains a saddle point \cite{Assyyaee:2015ale,paper4}.
This is an important criterion for having the structure of double field branes, since if we have only one minima at $\psi=0$, the brane reduces to the simple $\phi^{4}$ brane.

Note that for $\lambda=1$, $\alpha_{\phi}=1$ and $\beta^{2}=1$, since the vacua of the system reside at $(\phi, \psi)=(\pm 1, 0)$, only the $\phi$-field is responsible for the topological charge\footnote{In $1+1$ dimensions, the topological current is defined by $J^{\mu}=\frac{1}{2\pi}\partial^{\mu}\varphi$, from which $Q=\int_{-\infty}^{+\infty}J^{0}dx=\frac{1}{2\pi}[\phi(+\infty)-\phi(-\infty)]$  \cite{Guidry}.}(see Fig.\ref{2}) \cite{paper4}. In this situation, the symmetry of the system under $\phi\leftrightarrow -\phi$ and $\psi\leftrightarrow -\psi$ leads to the appearance of two similar branes with the same energy per unit surface. So, there are two types of kinks and antikinks which are related to each other by the field transformations $\phi\leftrightarrow -\phi$ and $\psi\leftrightarrow -\psi$ \cite{paper4}.

%%%%%%%%%%%%%%%%%%%%%%%%%%%%%%%%%%%%%%%%%%%%%%%%%%%%
\subsection{Soliton solutions}
%%%%%%%%%%%%%%%%%%%%%%%%%%%%%%%%%%%%%%%%%%%%%%%%%%%%

In particular cases, finding analytical solutions (kinks  or branes) for a specific system, such as the sine-Gordon or the $\phi^4$ systems, is possible. However, for other potentials including the system under 
consideration, analytical solutions cannot be found and one must resort to an initial assumption and use a 
numerical code which is able to minimize the energy of the system or make some algebraic 
simplification to find accurate (very good approximation) solutions (for instance, see 
\cite{Riazi:2002ty,paper4}).
Taking into account that for $\beta^{2}\neq\lambda\alpha_{\phi}^{2}$ there are two distinct degenerate vacua on the ($\phi$, $\psi$) plane, one expects combined soliton solutions for the ($\phi$, $\psi$) system. In the present context, the former (non-topological) case occurs when the $U(1)$ symmetry is not broken (left panel of Fig. \ref{2}).
From soliton theory, we know that in two field models in $1 + 1$ dimensions, we have topological solitons only if the degenerate vacua in the $\phi-\psi$ plane are disconnected (for more details, see \cite{Guidry}).

Approximate soliton solutions are given by \cite{Assyyaee:2015ale}:
\begin{eqnarray}\label{kak}
\phi(w)&=&\pm\alpha_{\phi}\tanh(\beta w), \nonumber\\
\psi(w)&=&\pm\alpha_{\psi} \sech(\beta w),
\end{eqnarray}
where
\begin{eqnarray}
\alpha_{\phi}=\alpha, \qquad \qquad
%\nonumber\\
\alpha_{\psi}=\sqrt{\alpha_{\phi}^{2}-2\frac{\beta^{2}}{\lambda}},
\end{eqnarray}
which correspond to the well-known exact topological and nontopological solutions in soliton theory. These solutions satisfy the static nonlinear field equations in the presence of the $U(1)$ symmetry system $\beta^{2}=0$ \cite{paper4} and are plotted in Fig. \ref{3}.
The positive (negative) sign of each field corresponds to a kink (antikink). On the other hand, $\alpha$ and $\beta$ are free parameters which control the height and the thickness of the brane, respectively.  More specifically, one can show that the height of the brane field is proportional to the $\alpha$ parameter while its thickness is given by $\triangle=\beta^{-1}$ \cite{Peyravi:2015bra}.

While double field models have been studied elsewhere (see e.g. \cite{Bazeia:1997zp}), the potentials in these references are different from the one considered here.  For instance, in \cite{Bazeia:1997zp} the solutions are obtained via minimizing the energy (Bogomolny bounds) and imposing the parity restrictions.
One can easily show that for the solutions (\ref{kak}), we have 
\begin{equation}
\frac{\phi^{2}}{\alpha_{\phi}^{2}}+\frac{\psi^{2}}{\alpha_{\psi}^{2}}=1.
\label{elipse}
\end{equation}
which represents an ellipse in the $(\phi, \psi)$ plane (see Fig.\ref{2}). In fact, this elliptic arc is responsible for connecting the two minima $(\pm 1, 0)$ of the corresponding potential (\ref{a}) \cite{Bazeia:1997zp}. It is worthwhile to note that while the one-field solutions demonstrate standard domain walls, the two-field solutions may represent domain walls with internal structure\footnote{The internal strucure shows itself in the apperance of asymetric shoulders of the brane, in Figs. \ref{t} and \ref{RR}}, which is clear by comparing the vector $(\phi,  \psi)$ configuration of one and two-fields solutions \cite{Bazeia:1997zp}. Note that this vector corresponds to a straight line sector and an elliptic arc for the one and two-field solutions, respectively \cite{Bazeia:1997zp}.

\begin{figure*}[ht!]
\centerline{\epsfxsize=16cm\epsfbox{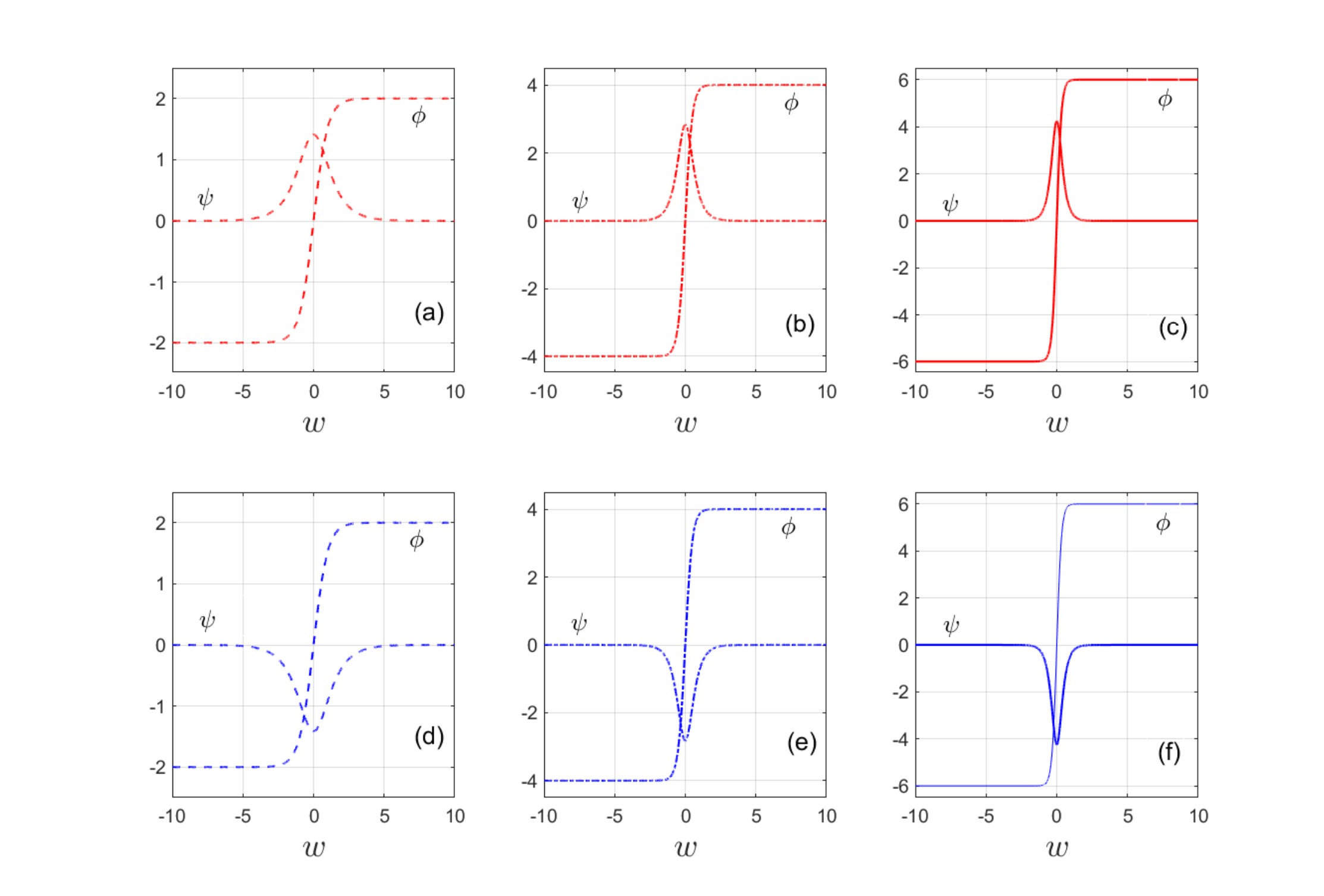}}
\caption{Soliton solutions as a function of the fifth dimension for the models with $\lambda=1$.
The plots (a), (b) and (c) depict the soliton \rom{1} ($\phi$, $\psi$) and (d), (e) and (f) the soliton \rom{2} ($\phi$, $-\psi$).
The dashed ($\phi(\alpha_{\phi}=2,\beta=1)$ and $\psi(\alpha_{\phi}=2,\beta=1)$), dotted-dashed ($\phi(\alpha_{\phi}=4,\beta=2)$ and $\psi(\alpha_{\phi}=4,\beta=2)$) and continuous curves ($\phi(\alpha_{\phi}=6,\beta=3)$ and $\psi(\alpha_{\phi}=6,\beta=3)$) correspond to solitons with decreasing brane thickness.\label{st}}
\end{figure*}

Each pair of these solutions are shown in Fig. \ref{3}.  It is seen that there are two types
of kinks and antikinks with the same energy which are related to each other by the field transformations $\phi\rightarrow-\phi$
and $\psi\rightarrow-\psi$. As one can see, under this operation, the soliton
$k_{1}$ (shown in Fig. \ref{3}(a)) changes to $\tilde{k}_{2}$ (shown in Fig. \ref{3}(d)) and $\tilde{k}_{1}$ (Fig. \ref{3}(c)) changes to $k_{2}$ (Fig. \ref{3}(b)).
In the rest of the paper, we will only consider $k_{1}$, the type \rom{1} solitons depicted in Fig. \ref{3}(a),  and $k_{2}$, the type \rom{2} solitons represented in Fig. \ref{3}(b). Figures \ref{st}(a) and \ref{st}(b) demonstrate soliton \rom{1} and soliton \rom{2} pairs for various values of parameters $\beta$, which correspond to branes with different thicknesses.

%\begin{figure*}[ht!]
%\epsfxsize=9cm\centerline{\hspace{8cm}\epsfbox{alphabetaM.pdf}\epsfxsize=9cm
%\centerline{\hspace{-10.2cm}\epsfbox{alphabetaManti.pdf}}}
%\caption{Soliton solutions as a function of the fifth dimension for the models with $\lambda=1$. 
%(a) for the soliton \rom{1} ($\phi$, $\psi$) and (b) for the soliton \rom{2} ($\phi$, $-\psi$).
%Dotted-dashed, dashed, and continuous curves correspond to solitons
%with decreasing brane thickness. \label{st}}
%\end{figure*}

%%%%%%%%%%%%%%%%%%%%%%%%%%%%%%%%%%%%%%%%%%%%%%%%%%%%
\subsection{Superpotentials}
%%%%%%%%%%%%%%%%%%%%%%%%%%%%%%%%%%%%%%%%%%%%%%%%%%%%

The formalism of our investigation is to keep the flat space soliton solution and modify the scalar
field potential in such a way that the soliton remains
a solution of the full gravitating system \cite{Peyravi:2015bra}. Thus, the soliton solution remains the same, while the form of the potential,
however, changes in such a way that the new set of equations with the brane geometry are satisfied. By plugging Eqs. (\ref{kak}) into Eqs. (\ref{8910b}) and (\ref{8910c}),
the superpotential of $\phi$ and $\psi$ (or $-\psi$) fields are given by:
\begin{eqnarray}
W_{1}(\phi)&=& 2\alpha_{\phi}\beta\left( \phi-{\phi^3\over{3\alpha_{\phi}^{2}}} \right),
\label{W1}
\end{eqnarray}
\begin{eqnarray}
W_{2}(\pm \psi)&=&\mp 2\alpha_{\psi}\beta\left[{\alpha_{\psi}\over 3}\left( 1 - {\psi^{2}\over \alpha_{\psi}^{2}}
\right)^{3/2}\right],
%W_{2}(-\psi)&=&+2\alpha_{\psi}\beta\left({\alpha_{\psi}\over 3}\left( 1 - {\psi^{2}\over \alpha_{\psi}^{2}}
%\right)^{3/2}\right).
\label{W2}
\end{eqnarray}
respectively.

Thus, one can define \cite{Dutra:2014xla}:
\begin{eqnarray}
W_{\rom{1}}(\phi,\psi)&=& W_{1}(\phi)+W_{2}(\psi),
	\nonumber\\
W_{\rom{2}}(\phi,-\psi)&=& W_{1}(\phi)+W_{2}(-\psi),
\end{eqnarray}
which are given as functions of $w$ through the following relations
\begin{eqnarray}
W_{\rom{1}}(w)&=&2\,{\alpha}^{2}\beta\,\tanh \left( \beta\,w \right) -\frac{4}{3}\,{\alpha}^{2}
\beta\, \left( \tanh \left( \beta\,w \right)  \right) ^{3}
	\nonumber \\
&&\qquad +\frac{4{\beta}^{3}}{3\lambda}\,{ \tanh \left( \beta\,w \right)^{3}},\\
W_{\rom{2}}(w)&=&2\,{\alpha}^{2}\beta\,\tanh \left( \beta\,w \right) -\frac{4{\beta}^{3}}{3\lambda}\,{ \tanh \left( \beta\,w \right)^{3}}.
\end{eqnarray}
Furthermore, taking into account condition (\ref{grant}), one can show that the solution curve in ($\phi$, $\psi$) plane is given by:
\begin{equation}\label{hs}
\phi(\psi)=\alpha_{\phi}\sqrt{1-\left( {\psi\over \alpha_{\psi}}\right)^{2}},
\end{equation}
which is a guarantee for the topological soliton. Note that Eqs. (\ref{elipse})--(\ref{W2}) and (\ref{hs}) impose a constraint ($|\psi|\leq|\alpha_{\psi}|$) on the $\psi$ field for the present solution.

Before examining the system precisely, one can predict similar results for the $\phi^{4}$ model due  to the presence of the topological soliton. However, because of the contribution from the second field, small deviations from the $\phi^{4}$ system are to be expected.

The corresponding modified potentials (\ref{mpot}) for this model are given by:
\begin{eqnarray}
&&V_{I,II}(\phi,\pm \psi)
= \frac{{\beta}^{2}}{2}\left[\,{\alpha_{{\phi}}}^{2}\left( 1-{\frac {{\varphi }^{2}}
{{\alpha_{{\phi}}}^{2}}} \right) ^{2}+ \left( 1-{\frac {{\psi}^{2}}{{\alpha_{{\psi}}}^{2}}} \right) {\psi}^{2}\right]
	\nonumber  \\
&& \quad -\frac{2}{3}\beta
\left[\alpha_{{\phi}}\left( \varphi -\frac{1}{3}\,{\frac {{\varphi}^{3}}{{\alpha_{{\phi}}}^{2}}}\right)
\mp \frac{1}{3}\,{\alpha_{{\psi}}}^{2}
\left( 1-{\frac {{\psi}^{2}}{{\alpha_{{\psi}}}^{2}}} \right)^{3/2} \right]^{2}.
\label{pot}
\end{eqnarray}
Note that while $\tilde{V}$ is of the order $O(\phi^{4})$ and $O(\psi^{4})$,
$V_{I,II}$ are $O(\phi^{6})$ and $O(\psi^{6})$. Furthermore, it is necessary to emphasize that in the limit of $\psi\rightarrow\alpha_{\psi}$ these potentials reduce to the $\phi^{4}$ potential \cite{Peyravi:2015bra}.

\begin{figure*}[ht!]
\epsfxsize=10cm\centerline{\hspace{10cm}\epsfbox{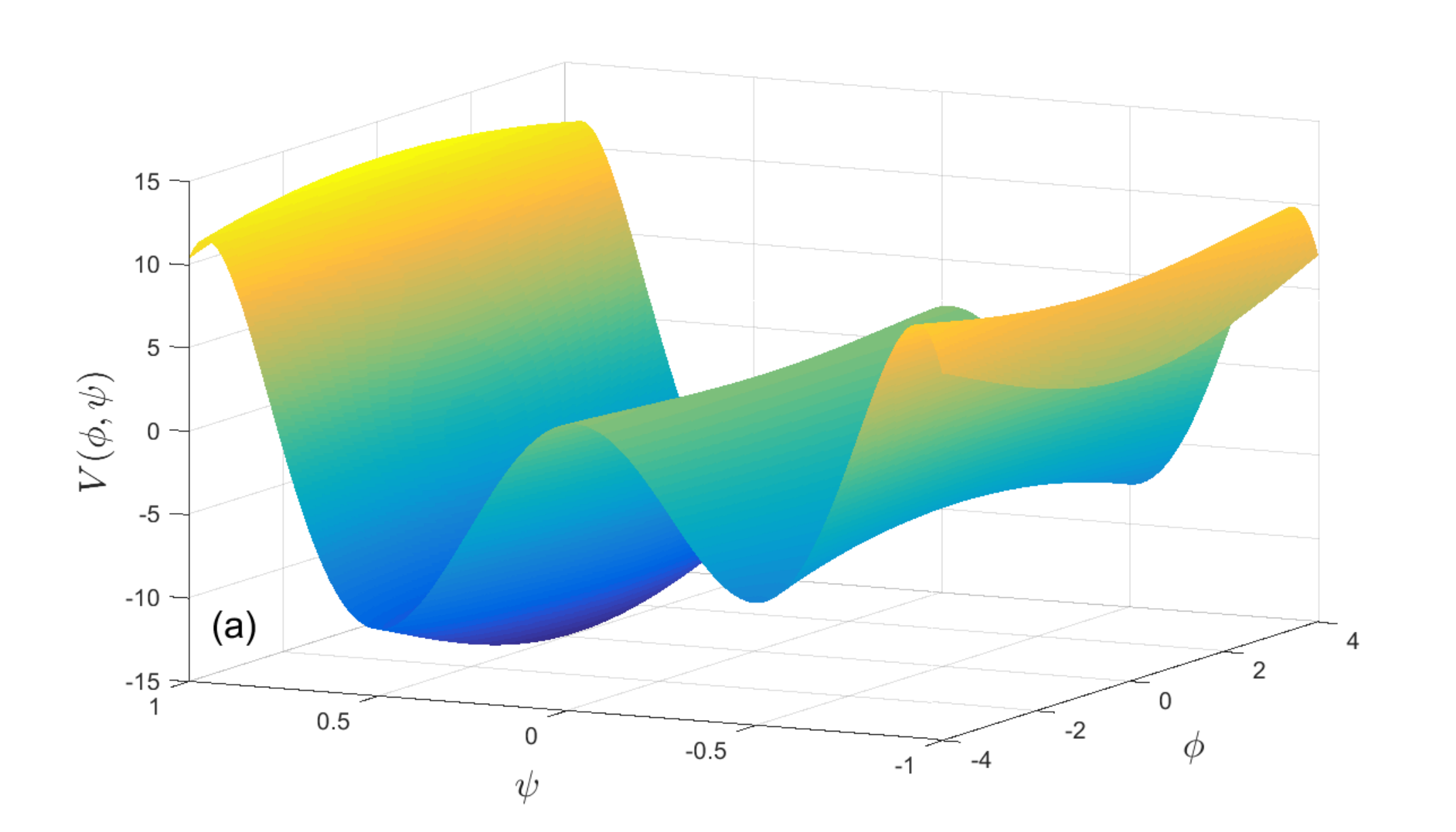}\epsfxsize=9cm\centerline{\hspace{-10.2cm}\epsfbox{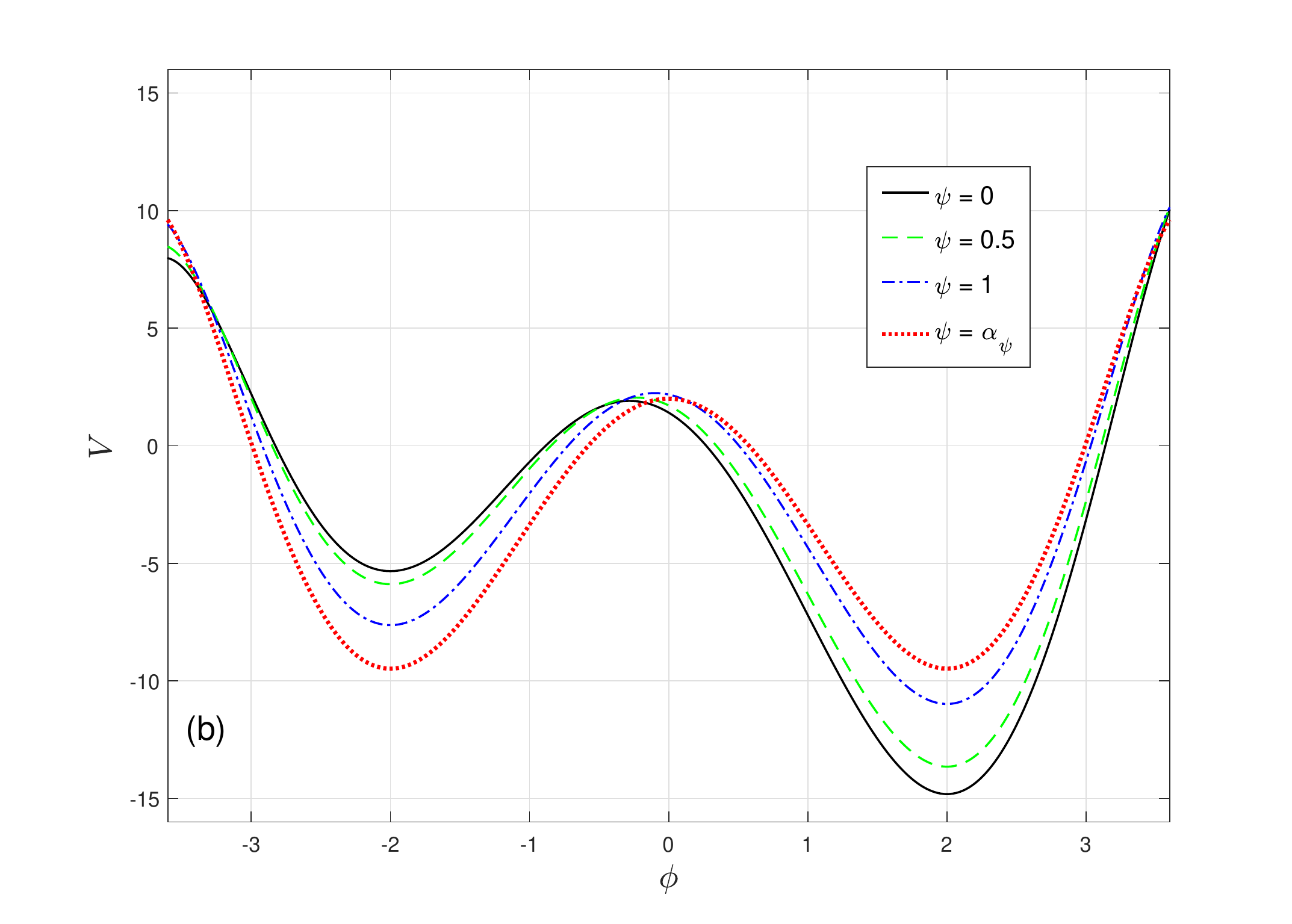}}}
\caption{The plots depict the modified soliton potential as a function of: (a) ($\phi$, $\psi$)  and (b)  $\phi$ for specific values of $\psi$. The potential has nondegenerate vacua in the $\phi$ direction for any values of $\psi$ except for $\lambda=1$, $\alpha_{\phi}=2$ and $\beta=1$. This behavior is that of a topological solitonic brane which does not have $Z_{2}$ symmetry.
\label{v}}
\end{figure*}

The potential (\ref{pot}) is plotted in Fig. \ref{v}. As the figure demonstrates,
nondegenerate vacua in the $\phi$ direction exist for any value of $\psi$ except for $\psi=\alpha_{\psi}$ (which corresponds to degenerate solutions type \rom{1} and \rom{2}) and this leads to
a topological solitonic brane with $Z_{2}$
symmetry breaking in the $\phi$ direction. Although this potential has been plotted for the soliton \rom{1}, it can be shown that for the soliton \rom{2} the general form of potential is unchanged.

%%%%%%%%%%%%%%%%%%%%%%%%%%%%%%%%%%%%%%%%%%%%%%%%%%%%
\subsection{Warp factors}
%%%%%%%%%%%%%%%%%%%%%%%%%%%%%%%%%%%%%%%%%%%%%%%%%%%%

The warp factors of system can be deduced from the field equations, are given by:
\begin{eqnarray}
&&\exp(2A_{\rom{1}}) = \left[ {\sech} ^{2}\left( \beta\,w \right)
 \right] ^{\frac{2}{9}{\alpha}^{2}+\frac{4}{9}{\frac {{\beta}^{2}}{\lambda}}}\times
	\nonumber  \\
&& \qquad \times \exp\left[\frac{4}{9}\frac{{\beta}^{2} \tanh ^{2}\left( \beta\,w
 \right)}{\lambda}-\frac{4}{9}{\alpha}^{2} \tanh^{2}\left( \beta\,w \right) \right],
 \label{warp1a}
\end{eqnarray}
\begin{eqnarray}
&& \exp(2A_{\rom{2}}) = \left[ {\sech}^{2} \left( \beta\,w \right)\right] ^{\frac{2}{3}{\alpha}^{2}-\frac{4}{9}{\frac {{\beta}^{2}}{\lambda}}} \times
	\nonumber  \\
&& \qquad \times
\exp\left[-\frac{4}{9}{\frac{{\beta}^{2} \tanh^{2}\left( \beta\,w\right)}{\lambda}}\right],
\end{eqnarray}
respectively, which in the limit $w\rightarrow\pm\infty$ are given by
\begin{eqnarray}
\exp(2A_{\rom{1}}) &\approx & 2\exp\left[ -2\beta\left(\frac{2}{9}\alpha^{2}+\frac{4}{9}\frac{\beta^{2}}{\lambda}\right)w\right]\times\nonumber\\
&&\exp\left[\frac{4}{9}\left(\frac{\beta^{2}}{\lambda}-\alpha^{2}\right)\right],
\end{eqnarray}
\begin{eqnarray}
\exp(2A_{\rom{2}}) &\approx & 2\exp\left[ -2\beta\left(\frac{2}{9}\alpha^{2}-\frac{4}{9}\frac{\beta^{2}}{\lambda}\right)w\right]\times\nonumber\\
&&\exp\left[-\frac{4}{9}\left(\frac{\beta^{2}}{\lambda}\right)\right].
 \label{warp2b}
\end{eqnarray}

\begin{figure*}[th!]
\includegraphics[scale=0.4085]{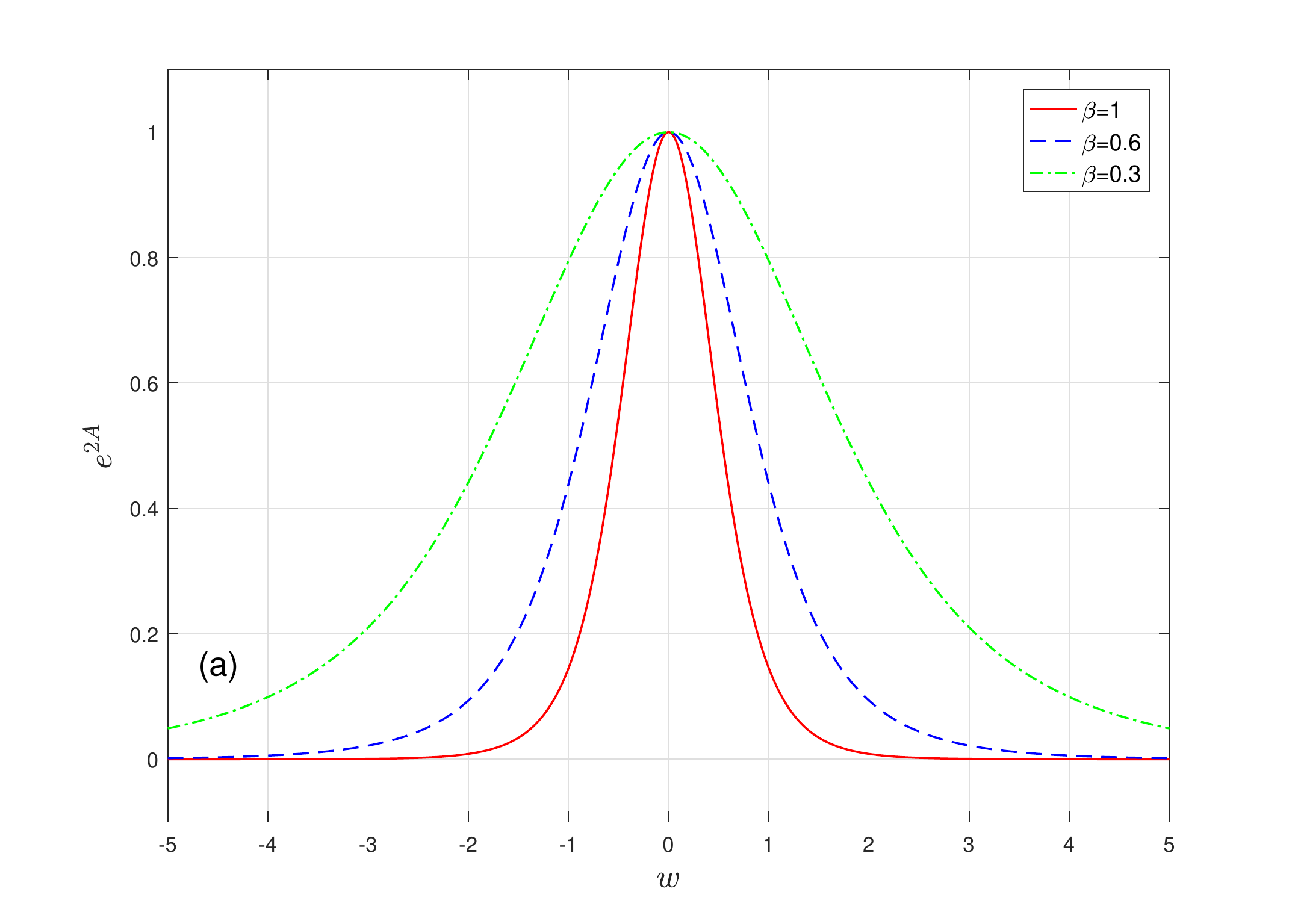}
\includegraphics[scale=0.4085]{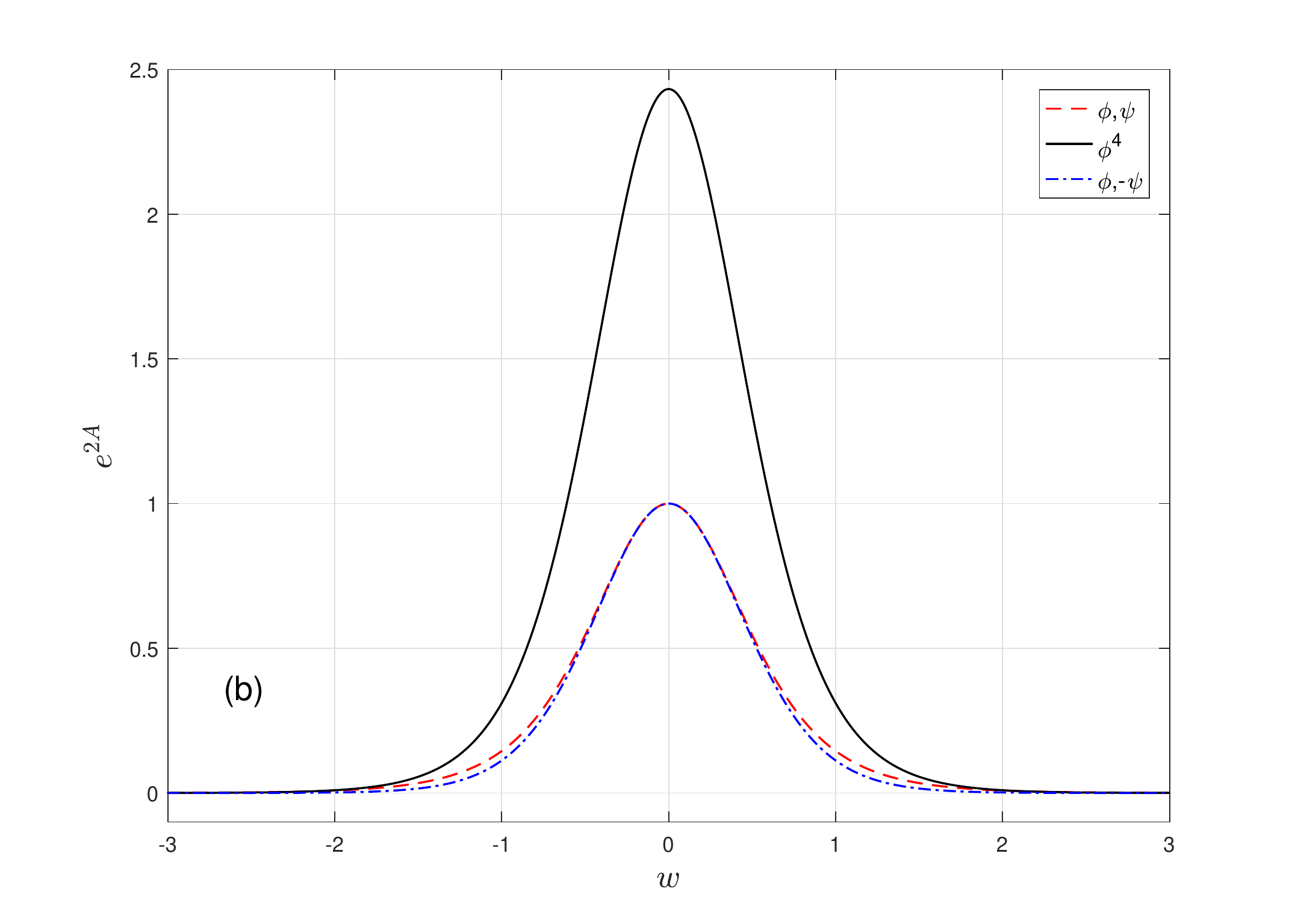}
\caption{The plots depict the warp factor as a function of the fifth dimension for the system. (a) For the soliton pair \rom{1} ($\phi$, $\psi$) for $\lambda=1$, $\alpha_{\phi}=2$ and different values of $\beta$. (b) The dashed, dotted-dashed and solid curves correspond to the warp factor of the soliton pair \rom{1} ($\phi$, $\psi$), soliton \rom{2} ($\phi$, $-\psi$) with $\lambda=1$, $\alpha_{\phi}=2$ and
$\beta=1$ and the $\phi^{4}$ model with  $\alpha=2$ and
$\beta=1$, respectively.}
\label{wp}
\end{figure*}

The warp factor is shown in Fig. \ref{wp}. Figure \ref{wp}(a) depicts the warp factor of the brane which is constructed by soliton \rom{1} for different values of the free parameters, while Fig. \ref{wp}(b) compares the warp factor of branes generated by type \rom{1},  type \rom{2} soliton pairs and the $\phi^{4}$ model, each for the same free parameters.
%As is transparent from the figure, the warp factor of the $\phi^{4}$ model is located between warp factors of the two considered solutions.

\begin{figure*}[th!]
\epsfxsize=9cm\centerline{\hspace{8cm}\epsfbox{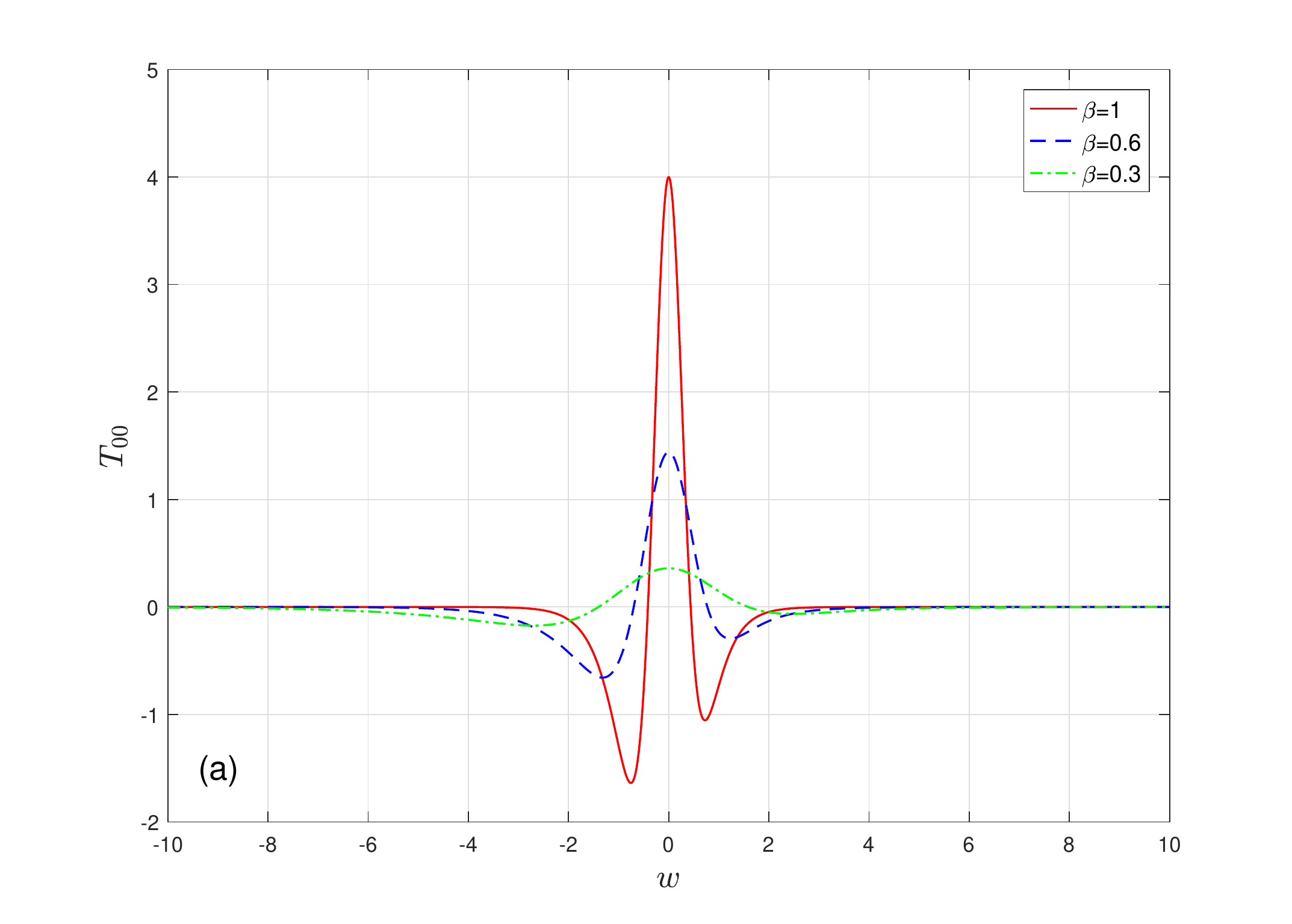}\epsfxsize=9cm\centerline{\hspace{-10.2cm}\epsfbox{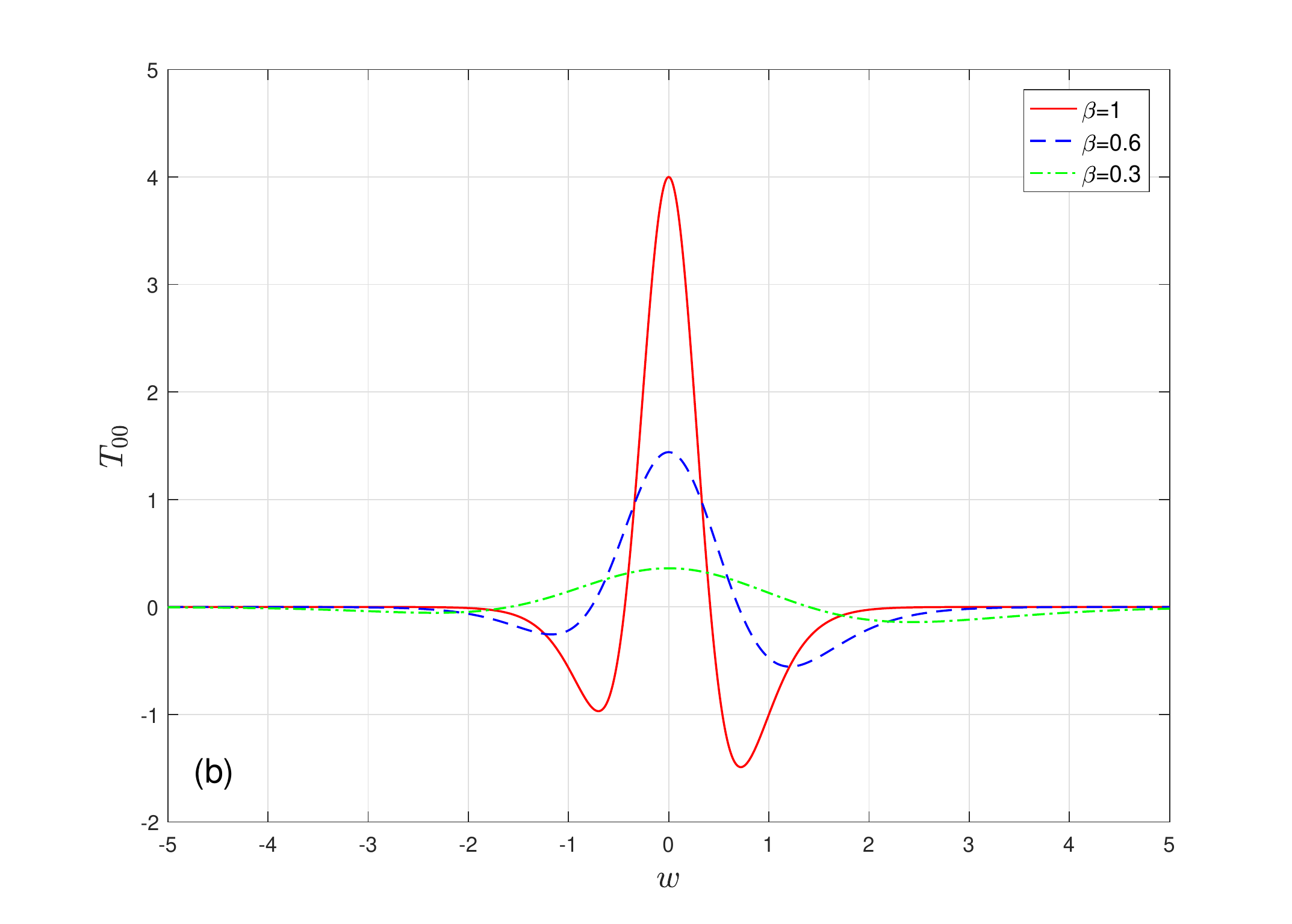}}}
\caption{The energy density (\ref{energydensity_blochb}) as a function of the fifth dimension for the system. For $\lambda=1$, $\alpha_{\phi}=2$ and different values of $\beta$, we have:
(a) the soliton pair \rom{1} ($\phi$, $\psi$) and (b) for the soliton pair \rom{2} ($\phi$, $-\psi$).\label{t}}
\end{figure*}

One can now analyse the energy density (\ref{energydensity_bloch}), which may be written as:
\begin{equation}
T_{00}=e^{2A}\left[\frac{1}{2}\phi'^{2}+\frac{1}{2}\psi'^{2}+V(\phi,\psi)\right].
\label{energydensity_blochb}
\end{equation}
As is evident from Fig. \ref{t} the energy density is localized, as expected. However, it contains two dips/shoulders on both sides of the brane, which are different for ($\phi$, $\psi$) and ($\phi$, $-\psi$) pairs, respectively.

%\begin{eqnarray}
%&&R_{\rom{1}}={\frac {320}{81}}\,{\frac {{\beta}^{2}}{{\lambda}^{2}}} \big( {\alpha}^{4}{\lambda}
%^{2}-2\,{\alpha}^{2}{\beta}^{2}\lambda + \,{\beta}^{4} \big)
%\tanh ^{6} \left( \beta\,w \right)
%	\nonumber  \\
%&&\quad +{\frac {32}{27}}\,{\frac {{\beta}^{2}}{{\lambda}^{2}}} \big( -10\,{\alpha}^{4}{\lambda}^{2}+10\,
%{\alpha}^{2}{\beta}^{2}\lambda -9\,{\alpha}^{2}{\lambda}^{2}+
%9\,{\beta}^{2}\lambda \big) \times
%	\nonumber \\
%&& \qquad  \qquad \qquad \times \tanh ^{4}\left( \beta\,w \right)
%	\nonumber  \\
%&& \quad +{\frac {16}{9}}\,{\frac {{\beta}^{2}}{{\lambda}^{2}}}
%\big( 9\,{\alpha}^{2}{\lambda}^{2}-6\,{\beta}^{2}\lambda
%+5\,{\alpha}^{4}{\lambda}^{2} \big) \tanh ^{2} \left( \beta\,w \right)
%\nonumber \\
%&& \qquad \qquad
% -\frac{16}{3}\,{\beta}^{2}{\alpha}^{2},
%\end{eqnarray}

%%%%%%%%%%%%%%%%%%%%%%%%%%%%%%%%%%%%%%%%%%%%%%%%%%%%
\subsection{Ricci and Kretschmann scalars}
%%%%%%%%%%%%%%%%%%%%%%%%%%%%%%%%%%%%%%%%%%%%%%%%%%%%

\begin{figure*}[ht!]
\epsfxsize=9cm\centerline{\hspace{8cm}\epsfbox{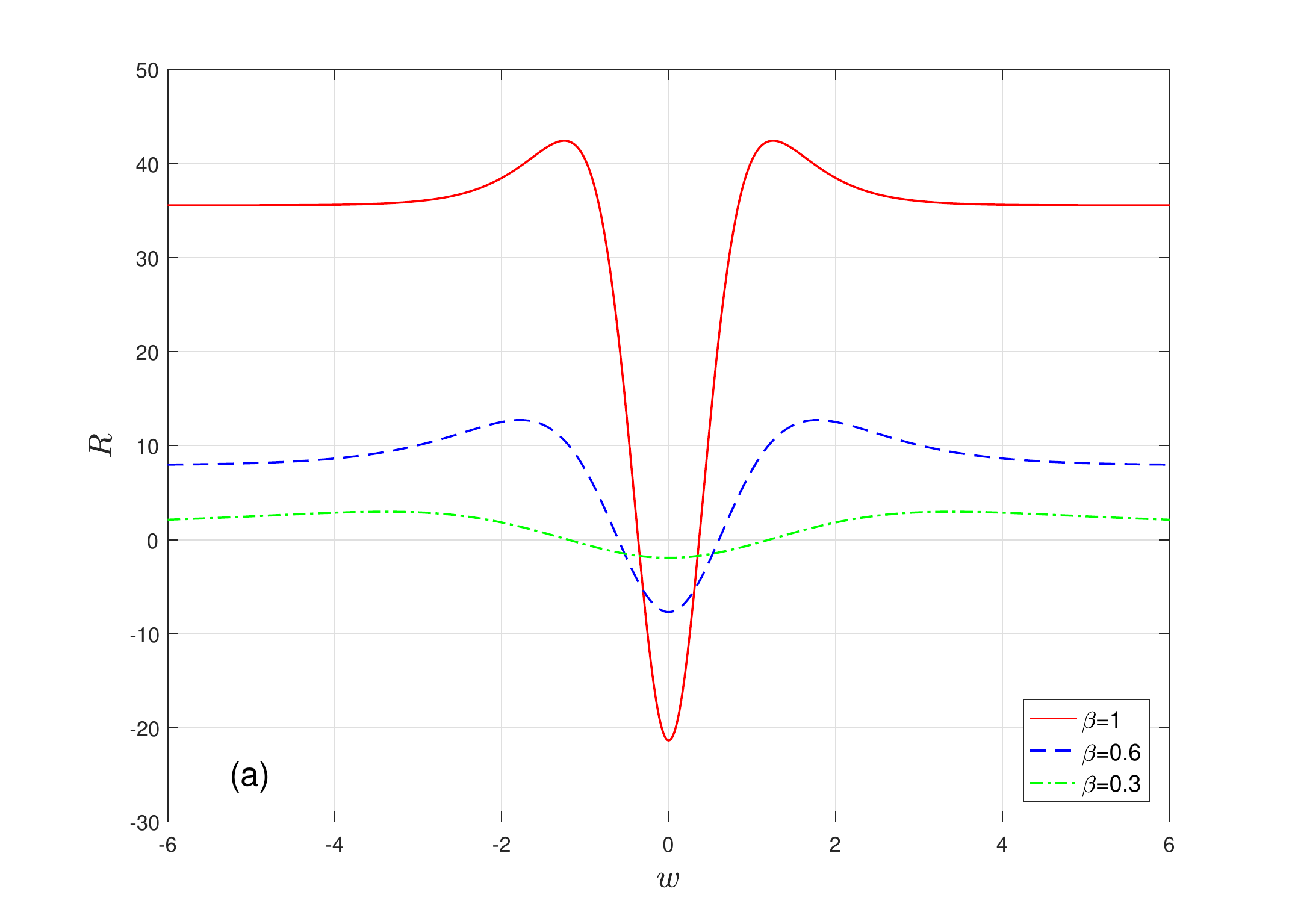}\epsfxsize=9cm\centerline{\hspace{-10.2cm}\epsfbox{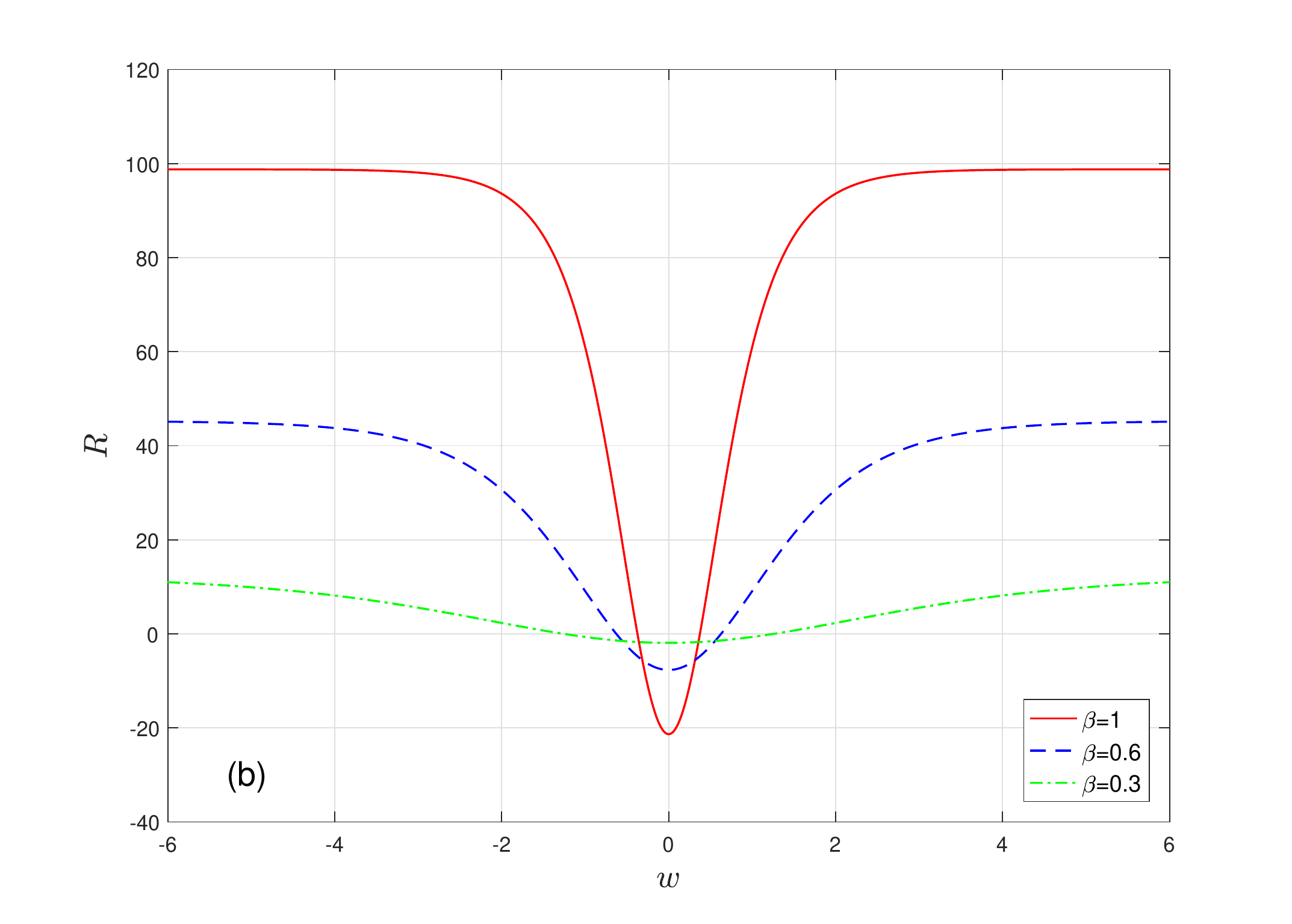}}}
\caption{Plots depict the Ricci scalar as a function of the fifth dimension for the system. For $\lambda=1$, $\alpha_{\phi}=2$ and different values of $\beta$.
(a) for the soliton pair \rom{1} ($\phi$, $\psi$) and (b) for the soliton pair \rom{2} ($\phi$, $-\psi$).\label{RR}}
\end{figure*}

\begin{figure*}[ht!]
\epsfxsize=9cm\centerline{\hspace{8cm}\epsfbox{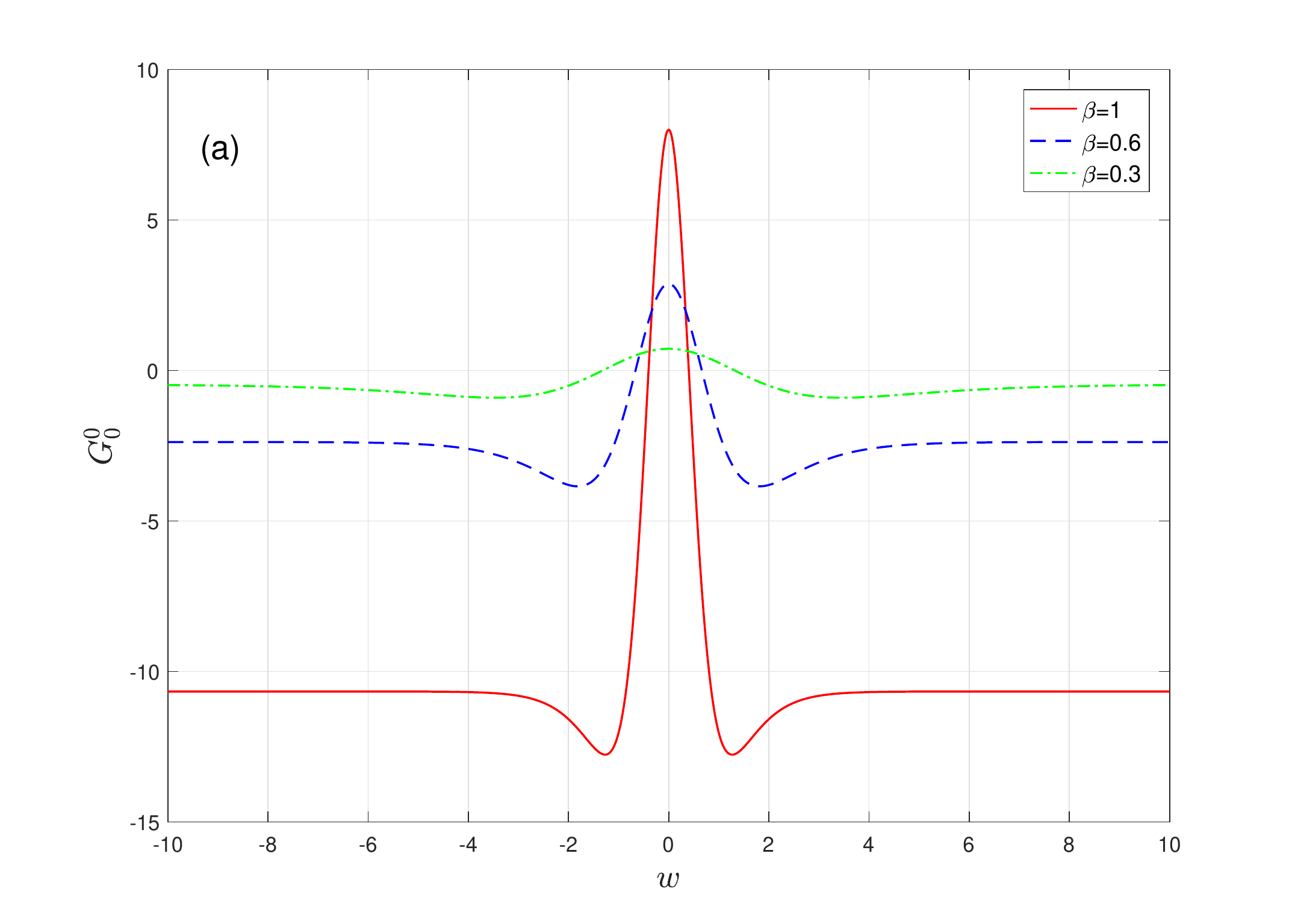}\epsfxsize=9cm\centerline{\hspace{-10.2cm}\epsfbox{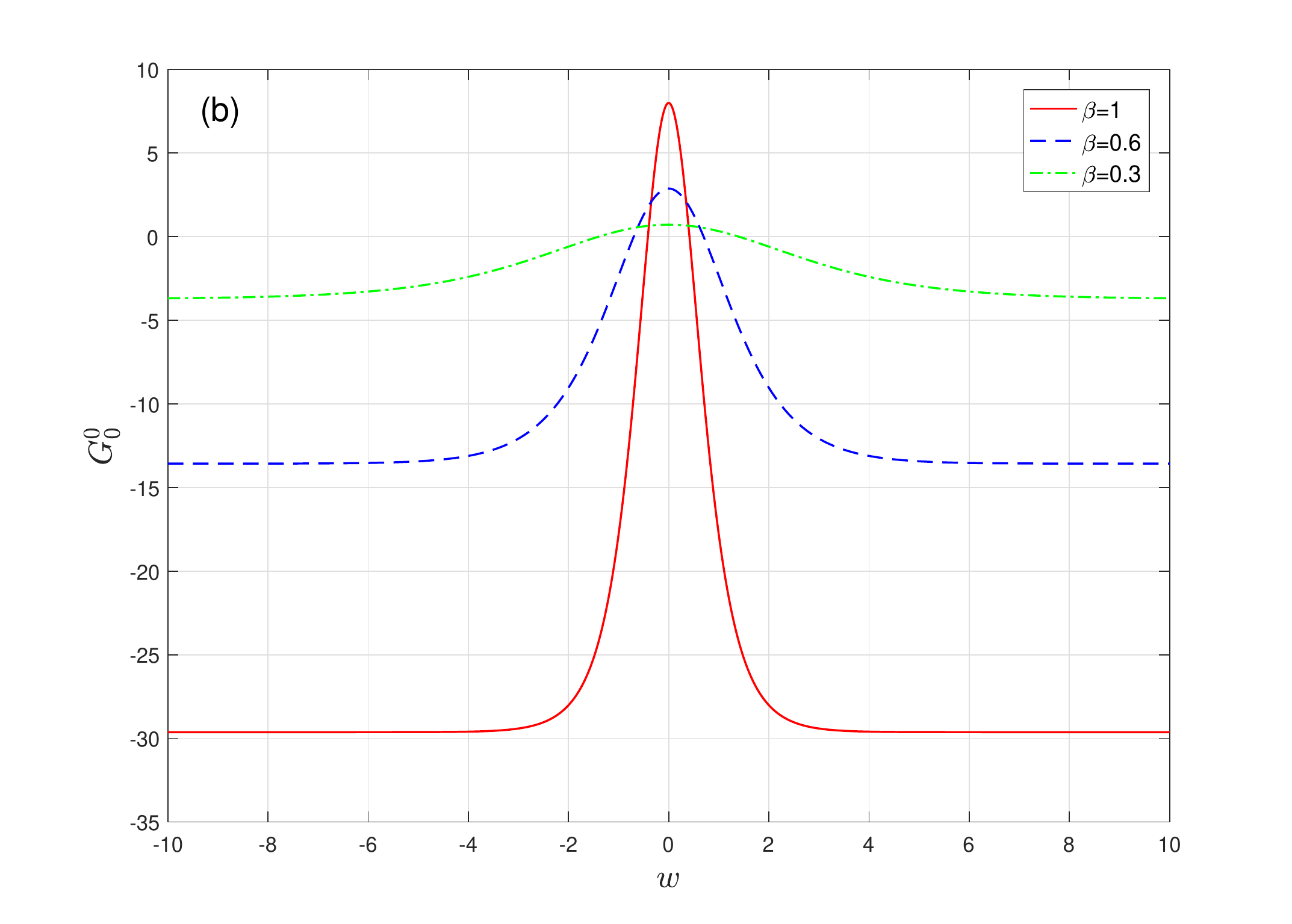}}}
\caption{Plots depict the Einstein tensor component $G^{0}_{0}$ as a function of the fifth dimension for the system. For $\lambda=1$, $\alpha_{\phi}=2$ and different values of $\beta$. (a) for the soliton pair \rom{1} ($\phi$, $\psi$) and (b) for the soliton pair \rom{2} ($\phi$, $-\psi$).\label{GG}}
\end{figure*}

It is useful to inspect the behaviour of the Ricci and Kretschmann scalars of the considered models. However, these are rather lengthy and are thus presented in Appendix \ref{Ricci_Kretschmann}, where it can be readily seen that there are no singularities. We do however show the behaviour of the Ricci scalar as a function of the fifth dimension $w$ in Fig. \ref{RR}, for the different soliton pairs ($\phi$, $\psi$) and ($\phi$, $-\psi$), respectively, and for different values of the free parameters.

In the limits of $w\longrightarrow\pm\infty$, the Ricci and Kretschmann scalars simplify to the following constant values:
\begin{equation}
\lim_{w\rightarrow\pm\infty} R_{\rom{1}}={\frac {1}{81}}\,{\frac {80\,{\beta}^{2}{\alpha}^{4}{\lambda}^{2}+320
\,{\beta}^{6}+320\,{\beta}^{4}{\alpha}^{2}\lambda}{{\lambda}^{2}}},
%	\nonumber \\
	\end{equation}
\begin{equation}
\lim_{w\rightarrow\pm\infty} R_{\rom{2}}={\frac {1}{81}}\,{\frac {320\,{\beta}^{6}-960\,{\beta}^{4}{\alpha}^{2}
\lambda+720\,{\beta}^{2}{\alpha}^{4}{\lambda}^{2}}{{\lambda}^{2}}}.
\end{equation}
and
\begin{eqnarray}
&&\lim_{w\rightarrow\pm\infty} K_{\rom{1}}={\frac {1}{6561\,\lambda^4}}\,\big(10240\,{\beta}^{12}+20480\,{\beta}^{10}{\alpha}^{2}\lambda
	\nonumber \\
&& \quad +15360\,{\beta}^{8}{\alpha}^{4}{\lambda}^{2}+5120\,{\beta}^{6}{\alpha}^{6}{\lambda}^{3}+
640\,{\beta}^{4}{\alpha}^{8}{\lambda}^{4} \big) ,\nonumber
\end{eqnarray}
\begin{eqnarray}
&&\lim_{w\rightarrow\pm\infty} K_{\rom{2}}=\frac {1}{6561\, \lambda^4}\,\big(-61440\,{\beta}^{10}{\alpha}^{2}\lambda + 10240\,{\beta}^{12} -
	\nonumber \\
&&  138240\,{\beta}^{6}{\alpha}^{6}{\lambda}^{3}+
138240\,{\beta}^{8}{\alpha}^{4}{\lambda}^{2}+51840\,{\beta}^{4}{\alpha
}^{8}{\lambda}^{4} \big),
\end{eqnarray}
respectively.

In the limit of $w\rightarrow0$, the Ricci scalars reduce to:
\begin{eqnarray}
\lim_{w\rightarrow0} R_{\rom{1}}&=&-\frac{16}{3}\,{\alpha}^{2}{\beta}^{2},
	\nonumber\\
\lim_{w\rightarrow0} R_{\rom{2}}&=&-\frac{16}{3}\,{\alpha}^{2}{\beta}^{2}.
\end{eqnarray}
and the Kretschmann scalars to:
\begin{eqnarray}
\lim_{w\rightarrow0} K_{\rom{1}}&=&{\frac {64}{9}}\,{\alpha}^{4}{\beta}^{4},
	\nonumber\\
\lim_{w\rightarrow0} K_{\rom{2}}&=&{\frac {64}{9}}\,{\alpha}^{4}{\beta}^{4},
\end{eqnarray}
respectively.

%%%%%%%%%%%%%%%%%%%%%%%%%%%%%%%%%%%%%%%%%%%%%%%%%%%%
\subsection{Mixed Einstein tensor components}
%%%%%%%%%%%%%%%%%%%%%%%%%%%%%%%%%%%%%%%%%%%%%%%%%%%%

The mixed Einstein tensor components of type \rom{1} and type \rom{2} branes are given by the equations in Appendix \ref{Einstein}, respectively, where we have $\mu = (1,2,3,4)$ as before.
As for the Ricci scalar, we show the behaviour of the mixed Einstein tensor component $G^{0}_{0}$ as a function of the fifth dimension $w$ in Fig. \ref{GG}, for the different soliton pairs ($\phi$, $\psi$) and ($\phi$, $-\psi$), respectively, and for different values of the free parameters. As can be seen, for the type \rom{1}, these diagrams are very similar to the $\phi^{4}$ model, while there is a small deviation from the $\phi^{4}$ model for the type \rom{2} soliton pair. It is worthwhile to notice that for both cases the bulk would be asymptotically anti-de Sitter on both side of these branes.

The components of the Einstein tensor for these two pairs in the limits $w\rightarrow\pm\infty$ become:
\begin{eqnarray}
\lim_{w\rightarrow\pm\infty} G^{A}_{B}&=&-\frac{8}{27}\frac{\beta^{2}}{\lambda^{2}}\big(\alpha^{4}\lambda^{2}+4\beta^{4}+4\beta^{2}\alpha^{2}\lambda\big)\nonumber,\nonumber\\
\lim_{w\rightarrow\pm\infty} G^{A}_{B}&=&{-\frac {32}{27}}\,{\frac {{\beta}^{6}}{{\lambda}^{2}}}+{\frac {32}{9}
}\,{\frac {{\beta}^{4}{\alpha}^{2}}{\lambda}}-\frac{8}{3}\,{\beta}^{2}{\alpha}^{4}.
\end{eqnarray}
and in the limit of $w\rightarrow0$ the Einstein tensor components take the form:
\begin{eqnarray}
\lim_{w\rightarrow0} G^{\mu}_{\nu}&=&-2\alpha^{2}\beta^{2}\delta^{\mu}_{\nu},
	\nonumber\\
\lim_{w\rightarrow0} G^{\mu}_{\nu}&=&2\alpha^{2}\beta^{2}\delta^{\mu}_{\nu},
\end{eqnarray}
respectively (we have assumed $8\pi G=2$). 
Note the cosmological constant on the brane would be $2\alpha^{2}\beta^{2}$ and $-2\alpha^{2}\beta^{2}$ for type {\rom 1} and type {\rom 2} respectively.
In this limit $G^{5}_{5}=0$ for both models, as it is expected. The cosmological constant on the
brane for the type {\rom 2} brane is exactly the same as in the $\phi^{4}$  model, while the cosmological constant of type {\rom 1} is the opposite \cite{Peyravi:2015bra}.

The linearized geodesic equation of a test particle moving in the direction of
the fifth dimension for both cases are given by:
\begin{equation}
\ddot{w}+c_{1}^{2}\frac{2}{3}\alpha^{2}\beta^{2}w=0\,,
\end{equation}
which corresponds to a frequency
$\omega=\Omega=\sqrt{F'(w_{0})}=\sqrt{2/3}\,c_{1}\alpha\beta$,
and is exactly the same as the frequency of the $\phi^{4}$ system.

%%%%%%%%%%%%%%%%%%%%%%%%%%%%%%%%%%%%%%%%%%%%%%%%%%%%
\subsection{Stability}\label{33}
%%%%%%%%%%%%%%%%%%%%%%%%%%%%%%%%%%%%%%%%%%%%%%%%%%%%

\begin{figure*}[ht!]
\epsfxsize=9cm\centerline{\hspace{8cm}\epsfbox{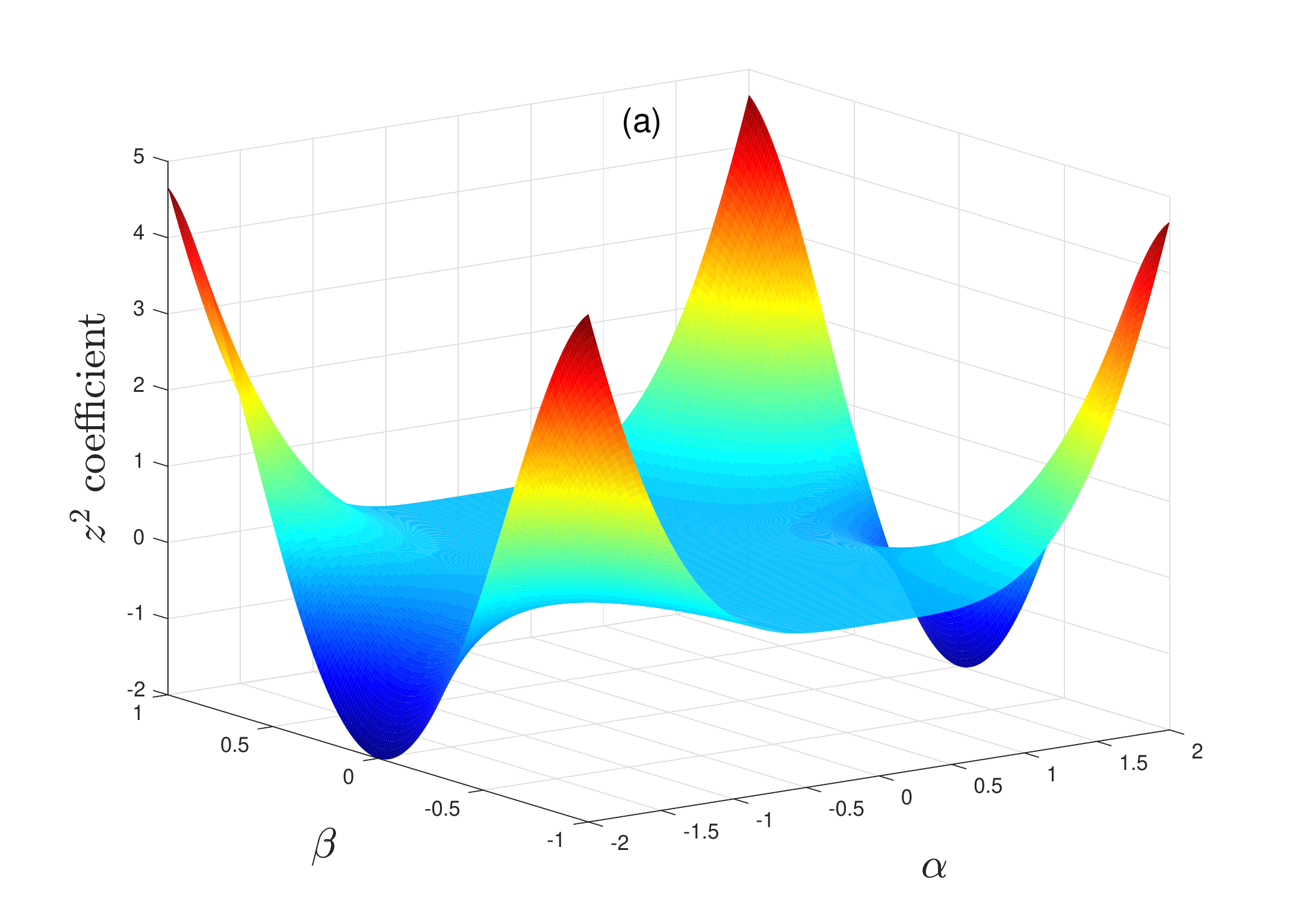}\epsfxsize=9cm\centerline{\hspace{-10.2cm}\epsfbox{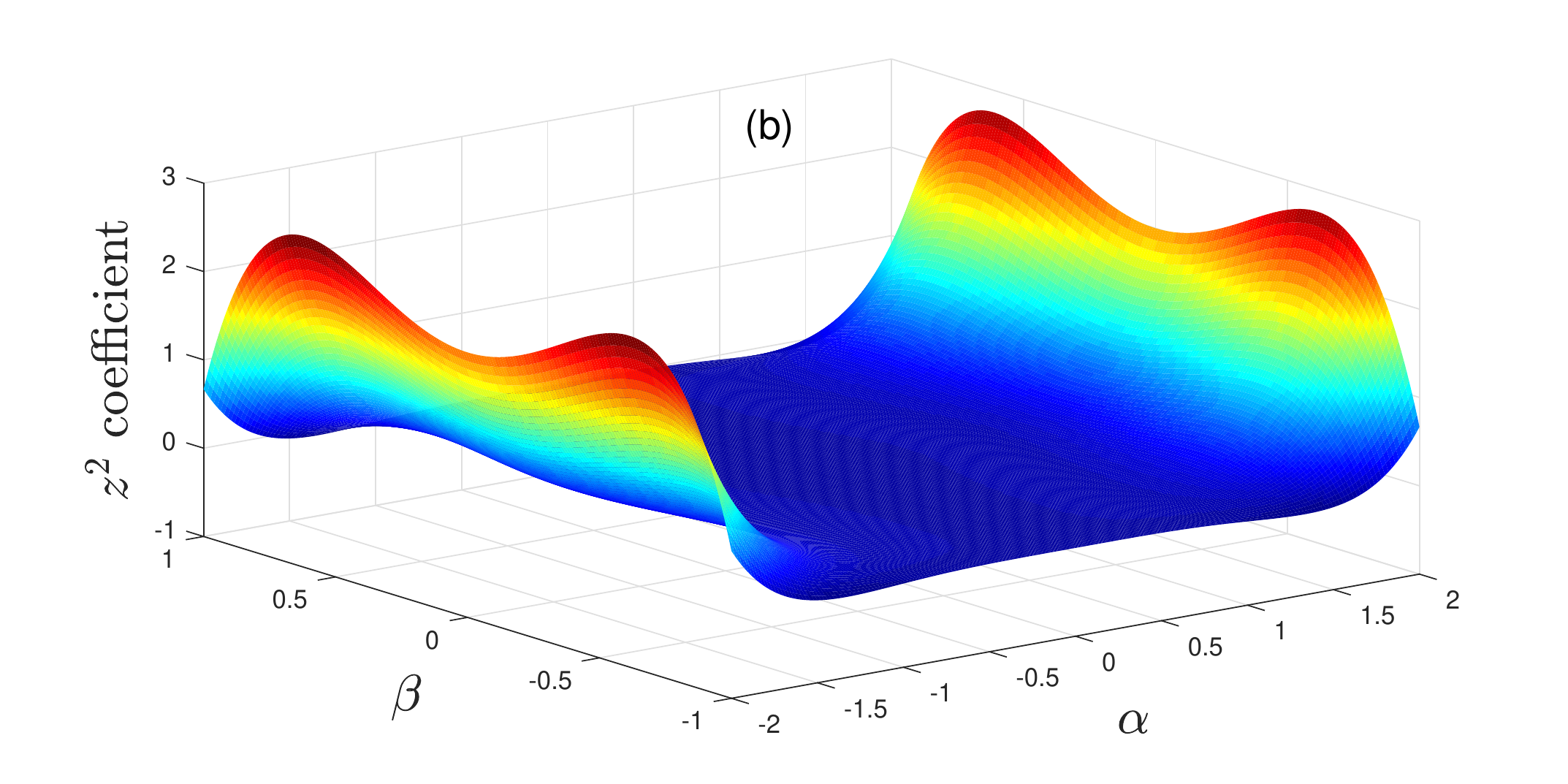}}}
\caption{Stability regions of the potential, where the plots depict the coefficient of the $z^2$ term in the potential of the linearised
Schr\"{o}dinger equation as a function of the free parameters
$\alpha$ and $\beta$ with $\lambda=1$ for the (a) pair of $\phi$ and $\psi$ (soliton \rom{1}) and (b) pair of $\phi$ and $-\psi$ (soliton \rom{2}).
Note that the sign of the $z^{2}$ term indicates the character of the stability, where the positive sign being stable, while the negative sign indicates instability.}
\label{U}
\end{figure*}

In order to study the stability of the branes, we choose an ``axial gauge'' in which the metric is perturbed in the following way \cite{DeWolfe:1999cp,Sadeghi:2007uk,Afonso:2006gi,Cruz:2014eca,Bazeia:2004dh}:
\begin{equation}
ds^{2}=e^{ 2A(w)}(g_{\mu\nu} + \varepsilon h_{\mu\nu} )dx^{\mu} dx^{\nu} - dw^{
2},
\end{equation}
where $g_{\mu\nu}$ is the four-dimensional metric,
$h_{\mu\nu}$ represents the metric perturbations, and $\varepsilon$ is a
small parameter \cite{Afonso:2006gi}. By considering $\phi\rightarrow\phi+\varepsilon\tilde{\phi}$ and $\psi\rightarrow\psi+\varepsilon\tilde{\psi}$ \cite{Bazeia:2004dh} and
variation of the action with respect to the scalar fields up to second order in
$\varepsilon$, one obtains the equations for the scalar fluctuations $\tilde{\phi}$ and $\tilde{\psi}$ as \cite{DeWolfe:1999cp,Bazeia:2004dh}:
\begin{eqnarray}
e^{-2A}\Box\tilde{\phi}-4A'\tilde{\phi}'-\tilde{\phi}''&+&\tilde{\phi}\frac{\partial^{2}V(\phi,\psi)}{\partial\phi^{2}}\nonumber\\
+\tilde{\phi}\frac{\partial^{2}V(\phi,\psi)}{\partial\phi\partial\psi}&=&\frac{1}{2}\phi'g^{\mu\nu}h'_{\mu\nu},\end{eqnarray}
and
\begin{eqnarray}
e^{-2A}\Box\tilde{\psi}-4A'\tilde{\psi}'-\tilde{\psi}''&+&\tilde{\psi}\frac{\partial^{2}V(\phi,\psi)}{\partial\psi^{2}}\nonumber\\
+\tilde{\psi}\frac{\partial^{2}V(\phi,\psi)}{\partial\psi\partial\phi}&=&\frac{1}{2} \psi' g^{\mu\nu}h'_{\mu\nu},
\end{eqnarray}
respectively\cite{DeWolfe:1999cp,Bazeia:2004dh}.

The variation of action with respect to the metric up to second order in $\varepsilon$ leads to \cite{DeWolfe:1999cp,Bazeia:2004dh}:
\begin{eqnarray}
&&-\frac{1}{2}\Box h_{\mu\nu} +e^{2A}\left(\frac{1}{2}\partial^{2}_{w}+2A'\partial_{w}\right)h_{\mu\nu}
-\frac{1}{2}g^{\lambda\rho}(\partial_{\mu}\partial_{\nu}h_{\lambda\rho}
	\nonumber\\
&& \qquad -\partial_{\mu}\partial_{\lambda}h_{\rho\nu}-\partial_{\nu}\partial_{\lambda}h_{\rho\mu})
%	\nonumber\\
+\frac{1}{2}g_{\mu\nu}e^{2A}A'\partial_{w}(g^{\lambda\rho}h_{\lambda\rho})
	\nonumber\\
&& \qquad \quad +\frac{4}{3}e^{2A}g_{\mu\nu}\left(\tilde{\phi}\frac{\partial V(\phi,\psi)}{\partial\phi}+\tilde{\psi}\frac{\partial V(\phi,\psi)}{\partial\psi}\right)=0.
\end{eqnarray}
Using the transformation $h_{\mu\nu}\rightarrow \bar{h}_{\mu\nu}=P_{\mu\nu\lambda\rho}h^{\lambda\rho}$ where \cite{Bazeia:2004dh}:
\begin{equation}
P_{\mu\nu\lambda\rho}=\frac{1}{2}(\pi_{\mu\lambda}\pi_{\rho\nu}+\pi_{\mu\rho}\pi_{\nu\lambda})-\frac{1}{3}\pi_{\mu\nu}\pi_{\lambda\rho},
\end{equation}
with the following definitions \cite{DeWolfe:1999cp,Bazeia:2004dh}:
\begin{equation}
\pi_{\mu\nu}=g_{\mu\nu}-\frac{\partial_{\mu}\partial_{\nu}}{\Box},
\end{equation}
\begin{equation}
\bar{h}''_{\mu\nu}+4A'\bar{h}'_{\mu\nu}=e^{-2A}\Box\bar{h}'_{\mu\nu},
\label{pedi}
\end{equation}
\begin{equation}
\bar{h}_{\mu\nu}=e^{ik.x}e^{\frac{-3}{2}A(z)}\chi_{\mu\nu}(z),
\end{equation}
where $\Box=g^{ij}\partial_{i}\partial_{j}$ in the denominator is the four-dimensional Laplacian resulting from nonlocal effects \cite{DeWolfe:1999cp}.

Moreover, in order to render the unperturbed metric conformally flat, one can choose $dz=e^{-A(w)} dw$. In this case, Eq. (\ref{pedi}) leads to the following Schr\"{o}dinger equation \cite{DeWolfe:1999cp,Sadeghi:2007uk,Afonso:2006gi,Bazeia:2004dh,Cruz:2014eca}:
\begin{equation}
-\frac{d^{2}\chi(z)}{dz^{2}}+U(z)\chi(z)=k^{2}\chi(z)\,,
\label{schrod}
\end{equation}
where the potential is given by:
\begin{equation}
U(z)=-\frac{9}{4}\Lambda+\frac{9}{4}A'^{2}+\frac{3}{2}A''.
\end{equation}
Note that $\Lambda$ is the cosmological constant on the brane, which could be positive, negative or zero corresponding to the $4D$ spacetime being de Sitter ($dS_{4}$),
anti-de Sitter ($AdS_{4}$) or Minkowski ($M_{4}$), respectively
\cite{Sadeghi:2007uk,Afonso:2006gi}.
Since the dependence of $U(z)$ on $z$ is fairly complicated, we limit ourselves to small $z$, which corresponds to the vicinity of the brane. If we expand $U(z)$ near to its minimum, the lowest order terms are a constant and a term quadratic in $z$. The sign of the $z^{2}$ term indicates the character of the stability. The positive sign being stable, while the negative sign indicates instability.

Figure \ref{U} shows the $z^2$ coefficient of the Taylor expansion of the linearised
Schr\"{o}dinger equation potential as a function of the free parameters
$\alpha$ and $\beta$. Comparing the two plots in the figure, which are related to distinct pairs, one observes that the type \rom{2} leads to stability, while the type \rom{1} involves neutral equilibrium for small values of $\alpha$ and $\beta$ and is unstable otherwise.

It is interesting to perform the localization of the graviton zero mode. To this effect, consider $k=0$ in Eq. (\ref{schrod}), which yields
\begin{equation}
-\dfrac{d^{2}\chi(z)}{dz^{2}}+U(z)\chi(z)=0.
\end{equation}
To find $U(z)$, we first have to deduce $A(z)$, for which we need $\omega(z)$ which is governed by $z=\int \exp{[-A(\omega)]}d\omega$. For neither of the solutions (\ref{warp1a})-(\ref{warp2b}), can we proceed analytically, even for the zero mode. However, if we model the potential with a quadratic one ($U(z)=U_{0}z^{2}$), the solutions will be the Bessel J and Y functions.
\begin{equation}
f(z)= C_{1}\sqrt{z}\; J\left(\dfrac{1}{4},\dfrac{\sqrt{C}}{2}z^{2}\right)+C_{2}\sqrt{z}\; Y\left(\dfrac{1}{4},\dfrac{\sqrt{C}}{2}z^{2}\right)\,,
\end{equation}
which yield wave-packet-like functions concentrated near $z=0$.

%%%%%%%%%%%%%%%%%%%%%%%%%%%%%%%%%%%%%%%%%%%%%%%%%%%%
\section{Conclusion}\label{44}
%%%%%%%%%%%%%%%%%%%%%%%%%%%%%%%%%%%%%%%%%%%%%%%%%%%%

Brane world scenarios are among popular cosmological models, in which the existence of extra dimensions leads to a possible solution to the hierarchy problem and explains the weakness of gravity compared to other forces in nature. In the orginal brane models, such as the Randall-Sundrum models \cite{Randall:1999ee,Randall:1999vf}, the brane is infinitely thin with respect to the extra dimension and standard model particles move only on the 3+1 dimensional brane. The appearance of a Dirac delta function with respect to the extra dimension is physically undesirable and in thick brane models, one tries to smooth out this singularity and replace the matching conditions at the brane boundary with the full gravitational equations. 
Most thick brane models are based on a real scalar field which is highly concentrated at the brane and rapidly tends to rest at its vacuum value in the bulk as we move from the brane location into the bulk. These models have a simple structure and ideas from soliton theory can be used to build such models. In double field models, such as the one worked out in this paper, one employs two scalar fields, instead of just one, to build thick branes with a richer internal structure. From soliton theory in $1+1$ dimensions, we know that topological solitons exist only when there are more than one minimum in the field potential and these minima are distinct and disconnected in the field space. 

Based on this imporant, yet simple concept, we have worked out a brane model, using two scalar fields which have initially a $U(1)$ symmetry, and then break this symmetry via an explicit term. In the symmetric mode, where the vacuum manifold is a circle ($S^1$), stable topological solitons do not exist, and therefore no stable thick brane model can be built. However, the insertion of the symmetry breaking term within the appropriate range of the symmetry breaking parameter, reduces the vacuum manifold to two distinct points along one of the scalar fields. This enables topological solitons to form, which have either positive or negative topological charges, depending on the vacuum field values chosen on either sides of the brane.
The main achievement of this paper consists in using the concept of an explicit symmetry breaking in building a stable two field thick brane model. We have rigorously examined various properties of the solutions, including particle motion across the brane, dynamical stability of the brane, and the behaviour of important geometrical quantities such as the Ricci and Kretschmann scalars. Our results show that the existence of a second field leads to a more structured brane with an asymptotically AdS bulk and a brane with positive cosmological constant.

\begin{widetext}

%%%%%%%%%%%%%%%%%%%%%%%%%%%%%%%%%%%%%%%%%%%%%%%%%%%%
\appendix
%%%%%%%%%%%%%%%%%%%%%%%%%%%%%%%%%%%%%%%%%%%%%%%%%%%%

%%%%%%%%%%%%%%%%%%%%%%%%%%%%%%%%%%%%%%%%%%%%%%%%%%%%
\section{Ricci and Kretschmann scalars}\label{Ricci_Kretschmann}
%%%%%%%%%%%%%%%%%%%%%%%%%%%%%%%%%%%%%%%%%%%%%%%%%%%%

The Ricci and Kretschmann scalars of the considered models are given by:

\begin{eqnarray}
R_{\rom{1}}&=&{\frac {16}{81}}\,{\frac {{\beta}^{2}}{{\lambda}^{2}}} \big( 20\,{\alpha}^{4}{\lambda}
^{2}-40\,{\alpha}^{2}{\beta}^{2}\lambda+20\,{\beta}^{4} \big)
\tanh ^{6} \left( \beta\,w \right) \cr
&&+{\frac {16}{81}}\,{\frac {{\beta}^{2}}{{\lambda}^{2}}} \big( -60\,{\alpha}^{4}{\lambda}
^{2}+60\,{\alpha}^{2}{\beta}^{2}\lambda-54\,{\alpha}^{2}{\lambda}^{2}+
54\,{\beta}^{2}\lambda \big) \tanh ^{4}\left( \beta\,w \right) \cr
&&+{\frac {16}{81}}\,{\frac {{\beta}^{2}}{{\lambda}^{2}}}
\big( 81\,{\alpha}^{2}{\lambda}^{2}-54\,{\beta}^{2}\lambda+45\,{
\alpha}^{4}{\lambda}^{2} \big) \tanh ^{2} \left( \beta\,w
\right) -\frac{16}{3}\,{\beta}^{2}{\alpha}^{2},
\end{eqnarray}
\begin{eqnarray}
R_{\rom{2}}&=&{\frac {320}{81}}\,{\frac {{\beta}^{6} \tanh^{6}\left( \beta\,w
\right)}{{\lambda}^{2}}}+{\frac {16}{81}}\,{\frac {{
\beta}^{2} \left( -54\,{\beta}^{2}\lambda-60\,{\alpha}^{2}{\beta}^{2}
\lambda \right) \tanh^{4} \left( \beta\,w \right)}{{
\lambda}^{2}}}\nonumber\\
&&+{\frac {16}{81}}\,{\frac {{\beta}^{2} \left( 27\,{
\alpha}^{2}{\lambda}^{2}+54\,{\beta}^{2}\lambda+45\,{\alpha}^{4}{
\lambda}^{2} \right)\tanh^{2} \left( \beta\,w \right) }{{\lambda}^{2}}}-\frac{16}{3}\,{\alpha}^{2}{\beta}^{2}.
\end{eqnarray}
and
\begin{eqnarray}
K_{\rom{1}} &=&{\frac {64}{6561}}\,{\frac {{\beta}^{4}}{{\lambda}^{4}}}\Bigl[\big( 160\,{\alpha}^{8}{
\lambda}^{4}+160\,{\beta}^{8}-640\,{\alpha}^{6}{\beta}^{2}{\lambda}^{3
}+960\,{\alpha}^{4}{\beta}^{4}{\lambda}^{2}-640\,{\alpha}^{2}{\beta}^{6}\lambda \big)\tanh^{12}\left(\beta\,w \right) \nonumber\\
&&+\big(-864\,{\alpha}^{6}{\lambda}^{4}+864\,{\beta}^{6}\lambda-960\,{\alpha}^{8}{
\lambda}^{4}+960\,{\alpha}^{2}{\beta}^{6}\lambda+2880\,{\alpha}^{6}{\beta}^{2}{\lambda}^{3}-2880\,{\alpha}^{4}{\beta}^{4}{\lambda}^{2} \nonumber\\
&& \qquad \qquad +2592\,{\alpha}^{4}{\beta}^{2}{\lambda}^{3}-2592\,{\alpha}^{2}{\beta}^{4}{\lambda}^{2}\big)\tanh^{10}\left(\beta\,w \right) \nonumber\\
&&+\big(-4320\,{\alpha}^{6}{\beta}^{2}{\lambda}^{3}+2160\,{\alpha}^{8
}{\lambda}^{4}+3888\,{\alpha}^{6}{\lambda}^{4}-864\,{\beta}^{6}\lambda
+2916\,{\alpha}^{4}{\lambda}^{4}+2916\,{\beta}^{4}{\lambda}^{2}-5832\,
{\alpha}^{2}{\beta}^{2}{\lambda}^{3} \nonumber\\
&& \qquad \qquad +2160\,{\alpha}^{4}{\beta}^{4}{\lambda}^{2}-8640\,{\alpha}^{4}{\beta}^{2}{\lambda}^{3}+5616\,{\alpha}
^{2}{\beta}^{4}{\lambda}^{2} \big)\tanh ^{8} \left( \beta\,w\right) \nonumber\\
&&+\big( -5832\,{\beta}^{4}{\lambda}^{2}-3024\,{\alpha}^{2}{
\beta}^{4}{\lambda}^{2}+14580\,{\alpha}^{2}{\beta}^{2}{\lambda}^{3}-
2160\,{\alpha}^{8}{\lambda}^{4}+9288\,{\alpha}^{4}{\beta}^{2}{\lambda}
^{3}+2160\,{\alpha}^{6}{\beta}^{2}{\lambda}^{3}-8748\,{\alpha}^{4}{
\lambda}^{4}\nonumber\\
&&  \qquad \qquad  -6264\,{\alpha}^{6}{\lambda}^{4} \big) \tanh^{6}
\left( \beta\,w \right)\nonumber \\
&&+\left( 9477\,{\alpha}^{4}{\lambda}^{4}-
3240\,{\alpha}^{4}{\beta}^{2}{\lambda}^{3}-11664\,{\alpha}^{2}{\beta}^
{2}{\lambda}^{3}+810\,{\alpha}^{8}{\lambda}^{4}+2916\,{\beta}^{4}{
\lambda}^{2}+4212\,{\alpha}^{6}{\lambda}^{4} \right) \tanh ^{4}
\left( \beta\,w \right)\nonumber\\
&&+\left( -4374\,{\alpha}^{4}{\lambda}^{4}-
972\,{\alpha}^{6}{\lambda}^{4}+2916\,{\alpha}^{2}{\beta}^{2}{\lambda}^
{3} \right) \tanh ^{2}\left( \beta\,w \right)\Big]+{\frac {64}{9}}\,{\beta}^{4}{\alpha}^{4},
\end{eqnarray}
\begin{eqnarray}
K_{\rom{2}} &=& {\frac {10240}{6561}}\,{\frac {{\beta}^{12}}{{\lambda}^{4}}}\tanh ^{12}\left( \beta
\,w \right)+{\frac {64}{6561}}\,{
\frac {{\beta}^{4}}{{\lambda}^{4}}}\Bigl[\big( -960\,{\alpha}^{2}{\beta}^{6}\lambda-864\,{
\beta}^{6}\lambda \big) \tanh ^{10}\left( \beta\,w \right) \cr
&&+ \big( 3024\,{\alpha}^{2}{\beta}^{4}{\lambda}^{2}+864\,{\beta}^{6}
\lambda+2916\,{\beta}^{4}{\lambda}^{2}+2160\,{\alpha}^{4}{\beta}^{4}{
\lambda}^{2} \big)\tanh^{8} \left( \beta\,w \right)\nonumber\\
&&+ \big( -5832\,{\beta}^{4}{\lambda}^{2}-2160\,{\alpha}^{6}{\beta}^{2}{\lambda}^
{3}-3024\,{\alpha}^{2}{\beta}^{4}{\lambda}^{2}-2916\,{\alpha}^{2}{
\beta}^{2}{\lambda}^{3}-3240\,{\alpha}^{4}{\beta}^{2}{\lambda}^{3}
\big) \tanh^{6} \left( \beta\,w \right)\nonumber\\
&&+\big( 2916\,{\beta}^{4}{\lambda}^{2}+3240\,{\alpha}^{4}{\beta}^{2}{\lambda}^{3}+729\,{
\alpha}^{4}{\lambda}^{4}+5832\,{\alpha}^{2}{\beta}^{2}{\lambda}^{3}+
972\,{\alpha}^{6}{\lambda}^{4}+810\,{\alpha}^{8}{\lambda}^{4} \big)
\tanh ^{4}\left( \beta\,w \right)\nonumber\\
&&+\big( -1458\,{\alpha}^{4}{\lambda}^{4}-972\,{\alpha}^{6}{\lambda}^{4}-2916\,{\alpha}^{2}{\beta}^
{2}{\lambda}^{3} \big)\tanh^{2} \left( \beta\,w \right)
\Bigr]+{\frac {64}{9}}\,{\beta}^{4}{\alpha}^{4}.
\end{eqnarray}
respectively. It can be seen that these quantities are singularity-free.

%%%%%%%%%%%%%%%%%%%%%%%%%%%%%%%%%%%%%%%%%%%%%%%%%%%%
\section{Einstein tensor components}\label{Einstein}
%%%%%%%%%%%%%%%%%%%%%%%%%%%%%%%%%%%%%%%%%%%%%%%%%%%%

The mixed Einstein tensor components of solion \rom{1} and soliton \rom{2} are given by following equations respectively (we have considered that $\mu = (1,2,3,4)$:
\begin{eqnarray}
G^{\mu}_{\mu}&=&-{\frac {2}{27}}\,{\frac {{\beta}^{2}}{{\lambda}^{2}}}\big( 16\,{\alpha}^{4}{\lambda}
^{2}-32\,{\alpha}^{2}{\beta}^{2}\lambda+16\,{\beta}^{4} \big)\tanh^{6}\left( \beta\,w \right) \nonumber\\
&& -{\frac {2}{27}}\,{\frac {{\beta}^{2}}{{\lambda}^{2}}}\big( -48\,{\alpha}^{4}{\lambda}^{2}+48\,{\alpha}^{2}{\beta}^{2}\lambda-54\,{\alpha}^{2}{\lambda}^{2}+
54\,{\beta}^{2}\lambda \big) \tanh^{4}\left( \beta\,w \right)\nonumber\\
&&-{\frac {2}{27}}\,{\frac {{\beta}^{2}}{{\lambda}^{2}}}\big( 81\,{\alpha}^{2}{\lambda}^{2}-54\,{\beta}^{2}\lambda+36\,{
\alpha}^{4}{\lambda}^{2} \big) \tanh^{2}\left( \beta\,w
\right)+2\,{\beta}^{2}{\alpha}^{2},
\end{eqnarray}
\begin{eqnarray}
G^{5}_{5}&=&-{\frac {8}{27}}\,{\frac {{\beta}^{2}}{{\lambda}^{2}}}\big( 4\,{\beta}^{4}+4\,{\alpha
}^{4}{\lambda}^{2}-8\,{\alpha}^{2}{\beta}^{2}\lambda \big)
\tanh^{6}\left( \beta\,w \right)\nonumber\\
&&-{\frac {8}{27}}\,{\frac {{\beta}^{2}}{{\lambda}^{2}}}\big( 12\,{\alpha}^{2}{\beta}^{2}\lambda-
12\,{\alpha}^{4}{\lambda}^{2} \big)\tanh^{4} \left( \beta\,w
\right)-\frac{8}{3}\,{\beta}^{2}{\alpha}^{4}
\tanh^{2}\left( \beta\,w \right).
\end{eqnarray}
and
\begin{eqnarray}
G^{\mu}_{\mu}&=&-{\frac {32}{27}}\,{\frac {{\beta}^{6}}{{\lambda}^{2}}}\tanh ^{6}\left( \beta\,w
\right)-{\frac {2}{27}}\,{\frac {{
\beta}^{2}}{{
\lambda}^{2}}} \big( -54\,{\beta}^{2}\lambda-48\,{\alpha}^{2}{\beta}^{2}
\lambda \big) \tanh^{4} \left( \beta\,w \right)\cr
&&-{\frac {2}{27}}\,{\frac {{\beta}^{2}}{{
\lambda}^{2}}} \big( 27\,{\alpha
}^{2}{\lambda}^{2}+54\,{\beta}^{2}\lambda+36\,{\alpha}^{4}{\lambda}^{2
} \big) \tanh^{2}\left( \beta\,w \right)+2\,{\alpha}^{2}{\beta}^{2},
\end{eqnarray}
\begin{eqnarray}
G^{5}_{5}&=& -{\frac {32}{27}}\,{\frac {{\beta}^{6}}{{\lambda}^{2}}} \tanh^{6} \left( \beta\,w
\right)+{\frac {32}{9}}\,{\frac {{
\beta}^{4}{\alpha}^{2}}{\lambda}} \tanh^{4} \left( \beta\,w \right) -\frac{8}{3}\,{\beta}^{2}{\alpha}^{4} \tanh^{2} \left( \beta
\,w \right),
\end{eqnarray}
respectively.
\end{widetext}

%%%%%%%%%%%%%%%%%%%%%%%%%%%%%%%%%%%%%%%%%%%%%%%%%%%%
\acknowledgments{NR acknowledges the support of Shahid Beheshti University Research Council. MP acknowledges the support of Ferdowsi University of Mashhad via the proposal No. 32361. FSNL acknowledges funding from the research grants No. UID/FIS/04434/2019, No. PTDC/FIS-OUT/29048/2017, and CEECIND/04057/2017.}
%%%%%%%%%%%%%%%%%%%%%%%%%%%%%%%%%%%%%%%%%%%%%%%%%%%%

%%%%%%%%%%%%%%%%%%%%%%%%%%%%%%%%%%%%%%%%%%%%%%%%%%%%

%%%%%%%%%%%%%%%%%%%%%%%%%%%%%%%%%%%%%%%%%%%%%%%%%%%%

%%%%%%%%%%%%%%%%%%%%%%%%%%%%%%%%%%%%%%%%%%%%%%%%%%%%
\end{document}